\documentclass{aa}
\usepackage[colorlinks=true,citecolor=blue,anchorcolor=blue,filecolor=blue,linkcolor=blue]{hyperref}
\usepackage{graphicx}
\usepackage{amssymb,amsmath}
\usepackage{latexsym}
\usepackage{mathtools}
\usepackage{epsfig}
\usepackage{natbib}
\usepackage{txfonts}
\usepackage{mathrsfs}
\begin{document}

   \title{Oblique rings from migrating exomoons: A possible origin for long-period exoplanets with enlarged radii}
   %\subtitle{}
   %\titlerunning{}
%
   \author{Melaine Saillenfest\inst{1}
          \and
          Sophia Sulis\inst{2}
          \and
          Paul Charpentier\inst{3}
          \and
          Alexandre Santerne\inst{2}
          }
   \authorrunning{Saillenfest et al.}
   \institute{IMCCE, Observatoire de Paris, PSL Research University, CNRS, Sorbonne Universit\'e, Universit\'e de Lille, 75014 Paris, France\\
             \email{melaine.saillenfest@obspm.fr}
             \and
             Universit{\'e} Aix Marseille, CNRS, CNES, LAM, Marseille, France
             \and
             Universit{\'e} de Toulouse, CNRS, IRAP, Toulouse, France
             }
   \date{Received 26 April 2023 / Accepted 7 June 2023}

%%%%%%%%%%%%%%%%%%%%%%%%%%%%%%%%%%%%%%%%%%%%%%%%%%%%%%%%%%%%%%%%%%%%%%%%%%%%%%%%%%%%%%%%%%%

  \abstract
  % context heading (optional)
  {The extremely low density of several long-period exoplanets in mature systems is still unexplained -- with HIP\,41378\,f being archetypical of this category. It has been proposed that such planets could actually have normal densities but be surrounded by a ring observed approximately face on, mimicking the transit depth of a puffy planet. This configuration would imply that the equator of the planet is nearly perpendicular to its orbit plane, which is at odds with the formation process of gas giants. Yet, in the context of the Solar System planets, it has recently been shown that after gigayears of evolution, the tidal migration of a moon can naturally lead to a very tilted planet with a ring.}
  % aims heading (mandatory)
  {As exomoons are expected to be ubiquitous around giant exoplanets, this mechanism may be responsible for the anomalous radii of some observed exoplanets. In preparation for the future discoveries of the \emph{PLATO} mission, we present a simple method for checking the plausibility of this mechanism for a given exoplanet.}
  % methods heading (mandatory)
  {Analytical formulas give the probability density function of the relevant precession harmonics of the planet. For each harmonic, simple criteria set the moon mass and other properties required for the mechanism to operate.}
  % results heading (mandatory)
  {We applied this methodology to HIP\,41378\,f, and we show that in order to reproduce the observed configuration, a hypothetical former moon should have had a moon-to-planet mass ratio of a few times $10^{-4}$ (i.e. roughly the mass of our Moon) and have migrated over a distance of a few planet's radii on a gigayear timescale. These orders of magnitude match the properties of moons expected to exist around gaseous exoplanets.}
  % conclusions heading (optional)
  {We conclude that the migration of a former moon is a viable formation pathway for the proposed ring and tilt of HIP\,41378\,f. This example strengthens the ring hypothesis and motivates its application to other promising targets.}
  

  \maketitle

\section{Introduction}\label{sec:intro}
The so-called super-puff exoplanets have moderate masses (typically $\lesssim 15$~$M_\oplus$) but surprisingly large radii ($\gtrsim 6$~$R_\oplus$), giving them extremely low bulk densities ($\lesssim 0.3$~g\,cm$^{-3}$; see e.g. \citealp{Lee-Chiang_2016}). Although relatively rare, super-puffs form a growing class of exoplanets. Among the puffiest exoplanets with the longest orbital periods, we can cite the iconic HIP\,41378\,f, Kepler-87\,c, Kepler-79\,d,  Kepler-177\,c, and Kepler-51\,b, c, and d. Super-puffs must be distinguished from inflated hot Jupiters, which show a correlation between stellar irradiation and radius inflation (see e.g. \citealp{Laughlin-etal_2011,Lopez-Fortney_2016}). This correlation indicates that hot Jupiters have extended atmospheres connected in some way to their close proximity to the star (see e.g. \citealp{Burrows-etal_2000,Chabrier-Baraffe_2007,Batygin-etal_2011,Grunblatt-etal_2016}). A similar conclusion can be reached for short-period sub-Neptunes \citep{Pu-Valencia_2017,Millholland_2019}, but not for distant super-puffs, because they have much cooler equilibrium temperatures and undergo negligible star-planet tidal dissipation.

Initiated by the preprint of \cite{Santerne-etal_2019}, the low density of exoplanet HIP\,41378\,f, in particular, immediately raised much discussion. HIP\,41378\,f is mature ($2.1^{+0.4}_{-0.3}$~Gyr; \citealp{Lund-etal_2019}) and has a long period ($542$~days) and low equilibrium temperature ($300$~K). Its low density ($0.09\pm0.02$~g\,cm$^{-3}$) puts this planet among the puffiest exoplanets known to date. Even though other super-puffs are known, most of them are likely young and/or have shorter periods \citep{Lee-Chiang_2016}. Instead of a radius inflation, \cite{Akinsanmi-etal_2020} propose that HIP\,41378\,f could be a standard Neptune-sized planet surrounded by an inclined opaque ring that would mimic the transit depth of an inflated planet. As no significant distortion is visible in the transit ingress and egress of HIP\,41378\,f, the hypothetical ring should be optically thick and seen roughly face on. This configuration would imply that the obliquity of the planet\footnote{Not to be confused with the stellar obliquity (i.e. the angle between the spin axis of the star and the orbit pole of a given planet). Throughout this article, the term obliquity is exclusively used for the planetary obliquity (i.e. the angle between the spin axis of the planet and its orbit pole).} is nearly $90^\circ$.

The ring hypothesis was investigated by \cite{Piro-Vissapragada_2020} for other super-puff exoplanets. Good candidates are Kepler-87\,c, Kepler-79\,d, and Kepler-177\,c, even though their moderate temperatures -- as that of HIP\,41378\,f -- do not allow for water ice to exist around them. Therefore, unlike Saturn's ring, their rings would need to be composed of porous rocky particles. According to the results of \cite{Piro-Vissapragada_2020}, HIP\,41378\,f is currently the best candidate for a ring. Its long period would protect a ring against destructive irradiation levels and a strong warp due to the stellar torque; it also results in negligible star-planet tidal dissipation, which means that no particular mechanism would be required for the planet to maintain a large obliquity\footnote{High-obliquity equilibrium states also exist for short-period planets \citep{Millholland-Laughlin_2019,Millholland-Spalding_2020}; however, because of tidal despinning and obliquity damping, their obliquity needs to be continuously forced through dynamical interactions involving several planets (see also \citealp{Su-Lai_2022a,Su-Lai_2022b}).}. The low eccentricity of HIP\,41378\,f also guarantees a small level of orbital perturbations for the ring particles.

In order to determine the planets' atmospheric properties and test the ring hypothesis, near-infrared transmission spectra have been acquired for Kepler-51\,b and d \citep{LibbyRoberts-etal_2020}, Kepler-79\,d \citep{Chachan-etal_2020}, and HIP\,41378\,f \citep{Alam-etal_2022}. These spectra ended up being featureless, ruling out clear, low-metallicity atmospheres. The ring hypothesis is therefore not contradicted for these planets, but flat spectra can also be produced by high-altitude hazes or high-metallicity atmospheres. In fact, convincing atmospheric models have been put forward for Kepler-51\,b and d, as well as Kepler-79\,d (see also \citealp{Wang-Dai_2019,Gao-Zhang_2020,Ohno-Tanaka_2021}). Interestingly, these models of extended atmospheres appear to be inapplicable to HIP\,41378\,f as it is too massive ($M=12\pm 3$~$M_\oplus$), too cold, and too old.

The question of the possible physical composition of HIP\,41378\,f was explicitly tackled by \cite{Belkovski-etal_2022}. The authors show that photoevaporation is not nearly enough to explain the extreme density disparity between planet~f and other planets in the system. Moreover, the observed mass and radius of HIP\,41378\,f would require an envelope-to-core mass fraction larger than $75\%$ together with a high entropy (e.g. produced by recent collisions). Such a massive envelope is unlikely from the perspective of planetary formation, as it would require runaway gas accretion to have started precisely during the dissipation of the gas disc, and planet HIP\,41378\,f may not be massive enough anyway to have triggered runaway accretion.

Hence, the ring hypothesis appears to be favoured for HIP\,41378\,f, and it may apply as well to a restricted number of other observed super-puffs. Tidal rings are confined below the Roche limit, very close to their host planets. As such, they are strongly coupled to the centrifugal bulge of the planets, and they directly materialise their equatorial planes. In order to produce a substantial increase in a planet's transit depth (i.e. a very noticeable super-puff), its ring must be oriented roughly in the sky plane. This means that the planet's spin axis must point roughly along the observer's direction; its obliquity is therefore $\varepsilon\approx 90^\circ$ as proposed by \cite{Akinsanmi-etal_2020}. Such an exotic configuration may seem questionable from a formation point of view. Because of the angular momentum acquired during gas accretion, gaseous planets are expected to form with low obliquities. The obliquities of the Solar System giant planets are therefore interpreted as strong tracers of their dynamical evolution, and much effort is put into understanding their origin (see e.g. \citealp{Tremaine_1991,Ward-Hamilton_2004,Hamilton-Ward_2004,Boue-etal_2009,Boue-Laskar_2010,Morbidelli-etal_2012,Vokrouhlicky-Nesvorny_2015,Rogoszinski-Hamilton_2020,Rogoszinski-Hamilton_2021,Saillenfest-etal_2020,Saillenfest-etal_2021a,Saillenfest-etal_2022,Salmon-Canup_2022,Rufu-Canup_2022,Wisdom-etal_2022}). In this context, the ring hypothesis for super-puffs would greatly benefit from an underlying mechanism that may be responsible for their unusual configuration. The existence of such a mechanism would not certify whether a given planet does possess a ring or not, but it would show whether known dynamical processes are able to (or are even likely to) produce the proposed configuration.

In the Solar System, a substantial tidal migration of moons has recently been observed to be at play around gaseous planets (see \citealp{Lainey-etal_2009,Lainey-etal_2017,Lainey-etal_2020,Jacobson_2022}) -- even though it involves mechanisms of energy dissipation that are vastly different from those responsible for the well known rapid migration of our Moon (see e.g. \citealp{Farhat-etal_2022}). These results have strong implications for the orbital dynamics of moons around gaseous planets, but also for the gigayear-timescale dynamics of planetary spin axes. Indeed, moons affect the spin-axis precession rate of planets in a way that is intimately related to their distance (see e.g. \citealp{Boue-Laskar_2006}). The migration of a moon is therefore accompanied with a variation in the planet's spin-axis precession rate. In turn, this variation can drive the planet into a so-called secular spin-orbit resonance, that is, a resonance between the planet's spin-axis precession and one harmonic of its orbital nodal precession. As a matter of fact, this kind of resonances abound in multi-planetary systems. Provided that a planet has a substantially massive migrating moon, it may therefore be guaranteed to encounter one of these resonances sooner or later during its evolution. Once captured in resonance, the still ongoing migration of the moon produces a gradual tilting of the planet's spin axis (unless, as for the Earth, resonances are so numerous that they overlap massively; see \citealp{Laskar-Robutel_1993,NerondeSurgy-Laskar_1997}). This phenomenon is probably responsible for the $27^\circ$ obliquity of Saturn \citep{Saillenfest-etal_2021a,Saillenfest-etal_2021b,Wisdom-etal_2022}, and it is predicted to happen to Jupiter in the future \citep{Saillenfest-etal_2020}. It may also have played a role in the tilting of Uranus \citep{Saillenfest-etal_2022}.

When the planet's obliquity reaches $\varepsilon\gtrsim 70^\circ$, however, regular moons are known to be unstable in some range of distance \citep{Tremaine-etal_2009}. Interestingly, the migration of a single moon makes the system converge to this unstable zone, putting a dramatic end to the tilting process (see \citealp{Saillenfest-Lari_2021,Saillenfest-etal_2022}). At this point, the moon may be ejected or be destructed below the planet's Roche limit, eventually forming a tidal disc of debris. In the latter case, the final state of the system is a ringed planet with very high obliquity. This final state recalls the exotic configuration proposed for super-puff exoplanets. It would therefore be valuable to determine whether this mechanism could apply to them and provide a plausible dynamical background to the ring hypothesis.

In this article, we aim to present a generic methodology to assess whether the migrating-moon mechanism can realistically produce a tilted ring around a given exoplanet. Even though the number of known distant super-puffs is small today, the future \emph{PLATO} mission \citep{Rauer-etal_2014,Rauer-etal_2016} will considerably increase our knowledge of the population of long-period exoplanets -- including their masses through an intensive radial-velocity follow-up. In this context, we need efficient methods for a routine characterisation of the newly discovered planets and identification of the most interesting targets for follow up. For this reason, we design our methodology to be applicable even if the minimum amount of information about the planetary system is available (masses, periods, and sky-plane inclinations).

The article is organised as follows. In Sect.~\ref{sec:basics}, we recall the basics of the tilting mechanism. In Sect.~\ref{sec:orbit}, we compute the probability density function of the dominant orbital precession frequencies of a planet, and we present an example of application to the super-puff exoplanet HIP\,41378\,f. From these results, we estimate in Sect.~\ref{sec:moon} the mass and migration rate that a moon around this planet would need in order to trigger the full tilting mechanism. In Sect.~\ref{sec:capture}, we check that the resonance is large enough to enable an adiabatic capture and tilting, and we illustrate this mechanism with numerical simulations. We then discuss our results in Sect.~\ref{sec:discussion} and conclude in Sect.~\ref{sec:conclusion}.

\section{Basic mechanism}\label{sec:basics}
As shown by \cite{Saillenfest-Lari_2021}, the tilting of a planet from a low obliquity $\varepsilon$ up to $\varepsilon\approx 90^\circ$ can be achieved on a gigayear timescale via the tidal migration of a moon. This process occurs through the adiabatic drift of the system along the centre of a secular spin-orbit resonance. In this section, we recall the physical quantities involved and the conditions required to trigger this process. 

We write $I$ the orbital inclination of the planet and $\Omega$ its longitude of ascending node. We decompose the inclination dynamics of the planet in a quasi-periodic series truncated to $N$ terms:
\begin{equation}\label{eq:zeta}
   \zeta = \sin\frac{I}{2}\exp(i\,\Omega) = \sum_{j=1}^{N}S_j\exp[i\,\phi_j(t)]\,,
\end{equation}
where $S_j$ is a positive real constant, and $\phi_j(t) = \nu_j\,t + \phi_j^{(0)}$ evolves linearly over time $t$ with frequency $\nu_j$. Resonance capture from a low obliquity is possible only for resonances with a harmonic having a negative frequency $\nu_j$ such that $|\nu_j|\geqslant p$, where
\begin{equation}\label{eq:p}
   p = \frac{3}{2}\frac{\mathcal{G}M_\star}{a^3(1-e^2)^{3/2}}\frac{J_2}{\omega\lambda}
\end{equation}
is the characteristic spin-axis precession rate of the planet. In this expression, $\mathcal{G}$ is the gravitational constant, $M_\star$ is the mass of the star, $a$ and $e$ are the semi-major axis and eccentricity of the planet on its orbit around the star, $J_2$ is the second zonal gravity coefficient of the planet, $\omega$ is its spin rate, and $\lambda$ is its normalised polar moment of inertia. The parameters $J_2$ and $\lambda$ must be defined through the same normalising radius $R$ (which is generally chosen as the equatorial radius of the planet).

\begin{figure}
   \includegraphics[width=\columnwidth]{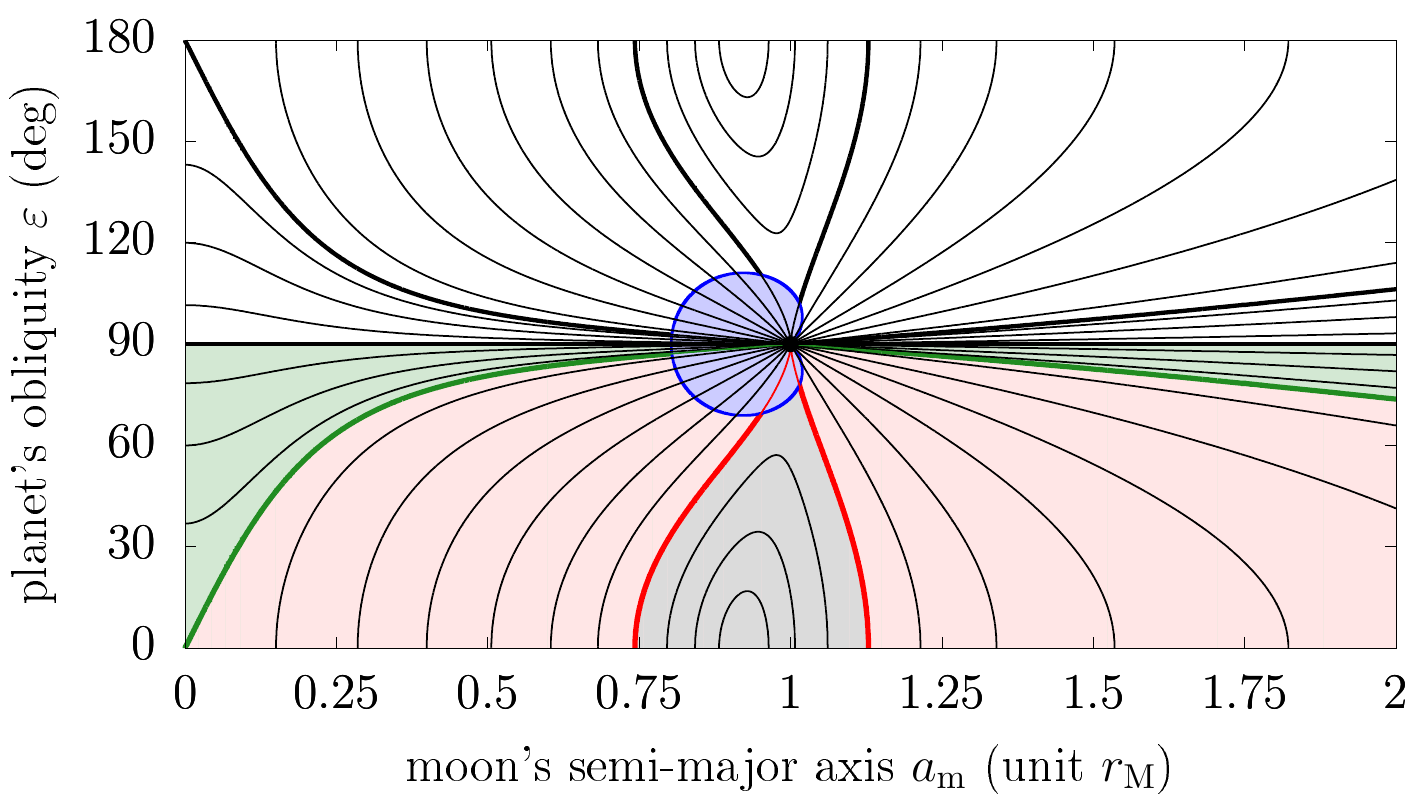}
   \caption{Level curves of the planet's spin-axis precession rate (adapted from \citealp{Saillenfest-etal_2022} for a moon mass-parameter $\eta=20$). If the planet is trapped in a secular spin-orbit resonance, the system evolves along one of these curves as the moon migrates. The condition in Eq.~\eqref{eq:condtilt} corresponds to the pink region; the left and right inequalities are the green and red curves, respectively. In the blue area, the moon is unstable.}
   \label{fig:precrate}
\end{figure}

The influence of a regular moon on the long-term spin-axis dynamics of the planet can be quantified by its non-dimensional `mass parameter' $\eta$ defined by
\begin{equation}
   \eta = \frac{1}{2}\frac{m}{M}\frac{r_\mathrm{M}^2}{J_2R^2}\,,
\end{equation}
where $m$ is the mass of the moon, $M$ is the mass of the planet, and $r_\mathrm{M}$ is the following characteristic length\footnote{$r_\mathrm{M}$ is called `mid-point radius' by \cite{Saillenfest-Lari_2021}. It is sometimes defined as the Laplace radius in other publications, either with or without the leading factor $2$.}:
\begin{equation}\label{eq:rM}
   r_\mathrm{M}^5 = 2\frac{M}{M_\star}J_2R^2~a^3(1-e^2)^{3/2}\,.
\end{equation}
Under the hypothesis that the moon's mass ratio $m/M$ is small (which does not necessarily imply that $\eta$ is small), \cite{Saillenfest-Lari_2021} show that all resonances with a nodal harmonic having a negative frequency $\nu_j$ verifying
\begin{equation}\label{eq:condtilt}
   p\leqslant|\nu_j|\leqslant p\frac{\eta}{2}
\end{equation}
can allow the planet's obliquity to grow from $\varepsilon=0^\circ$ to $\varepsilon=90^\circ$. This condition is illustrated in Fig.~\ref{fig:precrate}. Knowing the harmonics $\nu_j$ of the planet's orbital precession, Eq.~\eqref{eq:condtilt} allows one to compute the minimum mass required for the moon to produce the tilting. As resonances converge to an unstable region, the moon is ultimately lost at the end of the tilting process (see Fig.~\ref{fig:precrate}).

When Eq.~\eqref{eq:condtilt} is verified, the adiabatic capture and tilting of the planet within a given resonance requires an adequate hierarchy of timescales. First, we introduce the timescale $\tau$ of secular oscillations of the moon around its equilibrium `Laplace plane' (see \citealp{Tremaine-etal_2009}) as $\tau = 2\pi/\kappa$, where
\begin{equation}\label{eq:kappa}
   \kappa^2 = \frac{9}{4}\frac{M_\star}{M}\frac{ r_\mathrm{M}^3}{a^3(1-e^2)^3}\frac{\mathcal{G}M_\star}{a^3}\,.
\end{equation}
An adiabatic capture in resonance requires that $\tau$ is much shorter than the spin-axis precession timescale of the planet $T=2\pi/p$; this conditions is generally well verified in practice.

Then, a given observed planet may have been adiabatically tilted via a resonance only if the timescale $T_\mathrm{lib}$ of libration inside the resonance is much smaller than the age of the system. For a given secular spin-orbit resonance, the value of $T_\mathrm{lib}$ near the resonance centre can be computed as $T_\mathrm{lib} = 2\pi/\mu$, where
\begin{equation}\label{eq:Tlib}
   \mu^2 = (p')^2\left(\frac{\beta^2}{\sin^2\varepsilon_0} + \beta\sin\varepsilon_0\right)\,.
\end{equation}
In this expression, $\varepsilon_0$ is the planet's obliquity at the resonance centre and $p'$ is a modified version of $p$ that takes into account the presence of the planet's moon (see \citealp{Saillenfest-Lari_2021}). We define the non-dimensional variables $\gamma = \rho_1/p'$ and $\beta = \rho_2/p'$, where
\begin{equation}
   \begin{aligned}
      \rho_1 &= -\left(\nu_k - 2\sum_{j=1}^N\nu_jS_j^2\right)\,, \\
      \rho_2 &= -S_k\left(2\nu_k + \nu_kS_k^2 - 2\sum_{j=1}^N\nu_jS_j^2\right)\,, \\
   \end{aligned}
\end{equation}
and $k$ is the index in Eq.~\eqref{eq:zeta} of the considered resonance. $T_\mathrm{lib}$ depends on the distance of the moon through $p'$ and $\varepsilon_0$. However, an upper bound for $T_\mathrm{lib}$ is obtained at the time of resonance capture, for which $\gamma^{2/3} + \beta^{2/3} = 1$ (see \citealp{Henrard-Murigande_1987,Saillenfest-etal_2019}). In this case, $p'$ is equal to $p' = (\rho_1^{2/3} + \rho_2^{2/3})^{3/2}$, and the planet's obliquity at the centre of the resonance is
\begin{equation}
   \cos\varepsilon_0 = \gamma - \gamma^{1/3} + \sqrt{\gamma^2 + \gamma^{2/3} - \gamma^{4/3}}\,.
\end{equation}
Thanks to these expressions, we can compute $T_\mathrm{lib}$ from Eq.~\eqref{eq:Tlib} as a mere function of the planet's orbital dynamics in Eq.~\eqref{eq:zeta}.

\section{Orbital precession modes of the planet}\label{sec:orbit}

To apply this mechanism to a given planet, we need to know its orbital precession spectrum, which depends on planet-planet mutual interactions. However, the masses and orbital elements of exoplanets are generally not well known. For given parameters and their uncertainties, the most simple way to explore the variety of possible long-term orbital solutions is to use the Lagrange-Laplace system (see e.g. \citealp{Murray-Dermott_1999}).

\subsection{The Lagrange-Laplace proper modes}

The Lagrange-Laplace system is a secular theory at second order in eccentricity and inclination. As such, it  assumes that all eccentricities and inclinations are small and it neglects the long-term influence of mean-motion resonances. Small mutual inclinations are indeed strongly favoured in multi-planetary systems in which most planets are observed to transit their star. This is the case of HIP\,41378, around which the transits of five planets are observed \citep{Vanderburg-etal_2016}. Eccentricities are also expected to be small in multi-planetary systems for stability reasons. Moreover, according to the statistical distribution of multi-planetary systems \citep{Xie-etal_2016} and to theoretical arguments about chaotic diffusion (which leads to the statistical equipartition of angular momentum deficit; see \citealp{Laskar-Petit_2017}), planets having small mutual inclinations tend to have small eccentricities, and vice versa. Hence, the use of the Lagrange-Laplace theory is generally justified in this regard for multi-planetary systems. Neglecting the long-term effect of mean-motion resonances may seem more questionable, as many pairs of exoplanets are observed to be close to important resonances (see e.g. \citealp{Fabrycky-etal_2014}). Yet, the strongest mean-motion resonances in planetary systems -- and those enabling smooth captures -- are of eccentricity type. As such, they mainly affect eccentricities. Here, instead, we are only interested in the inclination degree of freedom of the planets because it is by far the main driver of their long-term spin-axis dynamics. The planets' eccentricity dynamics only enter into play at order three and beyond (see \citealp{Saillenfest-etal_2019}), so mean-motion resonances can safely be ignored in this analysis.

As above, we describe the nodal precession and inclination dynamics of a planet $k$ in the planetary system through the complex variable
\begin{equation}
   \zeta_k = \sin\frac{I_k}{2}\exp(i\,\Omega_k)\,,
\end{equation}
where $I_k$ is the orbital inclination of planet $k$ and $\Omega_k$ is its longitude of ascending node. The Lagrange-Laplace system gives the linear equation of motion
\begin{equation}\label{eq:laglap}
   \frac{\mathrm{d}\boldsymbol{\zeta}}{\mathrm{d}t} = iB\,\boldsymbol{\zeta}\,,
\end{equation}
in which $\boldsymbol{\zeta}$ is the vector containing the $\zeta_k$ variable of all planets and $B$ is a constant matrix that only depends on the masses and semi-major axes of the planets (see e.g. \citealp{Laskar-Robutel_1995}). The solution of this equation for a given planet $k$ has the form of a quasi-periodic series as in Eq.~\eqref{eq:zeta}:
\begin{equation}\label{eq:LLsol}
   \zeta_k(t) = \sum_{j=1}^{N_\mathrm{p}}S_j\exp\left[i\left(\nu_j\,t + \phi_j^{(0)}\right)\right]\,,
\end{equation}
where the number of terms $N$ is equal to the number of planets $N_\mathrm{p}$ in the system. Equation~\eqref{eq:LLsol} is a linear combination of proper modes whose frequencies $\nu_j$ are the eigenvalues of the matrix $B$. As $B$ only depends on the masses and semi-major axes of the planets, this is also the case of the frequencies $\nu_j$. Because of the conservation of total angular momentum, one of the frequencies $\nu_j$ is identically equal to zero; the related constant term in Eq.~\eqref{eq:LLsol} gives the orientation of the system's invariant plane.

Thanks to the fast computation of the solution of the Lagrange-Laplace system (which amounts to a mere matrix inversion), millions of trials can be performed at virtually no cost. In order to explore the distribution of possible values for the frequencies $\nu_j$, the first step is to draw the masses and semi-major axes of the $N_\mathrm{p}$ planets from their respective statistical distributions -- which represent our knowledge of their values. A similar approach was followed by \cite{Becker-Adams_2016} in their study of the compact multi-planetary systems observed by \emph{Kepler}. Each sequence of masses and semi-major axes for the $N_\mathrm{p}$ planets represent a possible realisation of the planetary system. In case the mass of a given planet has not been measured, a broad distribution of mass can be adopted (e.g. a uniform distribution in a given interval, or a law drawn for an assumed mass-radius relationship; see below). From a large number of realisations of the planetary system, a histogram for each frequency $\nu_j$ can be computed. These histograms define the possible locations of secular spin-orbit resonances given our current knowledge of the planetary system.

In practice, the largest values of the Lagrange-Laplace matrix $B$ in Eq.~\eqref{eq:laglap} are often located along its diagonal (meaning that the planetary system is only weakly coupled); this implies that each planet $k$ has its own dominant proper mode, which appears in Eq.~\eqref{eq:LLsol} as the term with largest amplitude. The frequency of the dominant proper mode of planet $k$ is usually noted $s_k$. In the context of the Lagrange-Laplace approximation, the quasi-periodic series in Eq.~\eqref{eq:LLsol} contains exactly $N=N_\mathrm{p}$ terms and the frequencies $\nu_j$ are each equal to one of the $s_k$. More generally, the orbital evolution of any planet in a stable system can be written as in Eq.~\eqref{eq:LLsol}, but where $N$ tends to infinity and each harmonic $\nu_j$ is a linear combination of the fundamental frequencies of the system  (see Sect.~\ref{sec:basics}). The first few strongest harmonics of the series are however proper modes given by the Lagrange-Laplace approximation; hence, the analysis presented here can be thought of as the dominant component of a more general theory.

While building the histogram for each proper mode $s_k$, a complication may arise. Indeed, if the masses and semi-major axes of the planets have large uncertainties, the distributions of the various frequencies may overlap. In this case, identifying each eigenvalue $\nu_j$ of matrix $B$ as the correct proper mode $s_k$ requires some caution. As the hierarchy of proper modes depends on the planetary system considered, a specific identification process is required. As an example, we subsequently present the case of the HIP\,41378 system.

\subsection{Application to the HIP\,41378 system}

HIP\,41378 is bright F-type star\footnote{Also known as K2-93 and EPIC~211311380.} which harbours at least five planets called b, c, d, e, and f \citep{Vanderburg-etal_2016}. Dynamical analysis reveals that planets~b and c are slightly off the 2:1 mean-motion resonance, similarly to many \emph{Kepler} planets (see e.g. \citealp{Fabrycky-etal_2014}). A tentative detection of a sixth planet -- planet~g -- is reported in the preprint of \cite{Santerne-etal_2019} close to the 2:1 mean-motion resonance with planet~c. As of today, only planets~b, c, and f have been observed during successive transits (see \citealp{Vanderburg-etal_2016,Becker-etal_2019,Bryant-etal_2021,Alam-etal_2022}) and unambiguously detected in radial velocity \citep{Santerne-etal_2019}. Therefore, only planets~b, c, and f have secured periods and masses.

Two transits of planet~d have been observed by the \emph{K2} mission but they are separated by a three-year observation gap, leading to a discrete set of possible periods \citep{Becker-etal_2019}. From stability considerations, and thanks to additional observations by \emph{TESS}, this discrete set is further reduced to only two likely values ($278$ and $371$~days; see \citealp{Berardo-etal_2019,Lund-etal_2019,Grouffal-etal_2022}). In contrast, only one transit of planet~e has been observed so far, so its period suffers from large uncertainties. The best period estimate for planet~e is $260^{+160}_{-60}$~days \citep{Lund-etal_2019}. The period of $369\pm 10$~days obtained by \cite{Santerne-etal_2019} is compatible with this estimate, and it results in a mass of $12\pm 5$~$M_\oplus$ for planet~e. The mass of planet~d, however, is unknown.

As explained in Sect.~\ref{sec:intro}, planet HIP\,41378\,f is a paradigmatic case of distant super-puff. Its period is about $542$~days, and it has a radius of $9.2\pm 0.1$~$R_\oplus$ and mass $12\pm 3$~$M_\oplus$, giving it a bulk density of $0.09\pm 0.02$~g\,cm$^{-3}$ \citep{Santerne-etal_2019}. Under the ring hypothesis, current data suggests a planet with radius $R=3.7\pm0.3$~$R_\oplus$ surrounded by a ring with radius $2.6\pm0.2$~$R$ and inclination $25\pm 4^\circ$ from the sky plane \citep{Akinsanmi-etal_2020}. This new planetary radius yields a bulk planet density of $1.2\pm 0.4$~g\,cm$^{-3}$, similar to that of Uranus. The hypothetical equatorial ring provides an indirect measure of the obliquity of the planet, namely\footnote{The ring obtained by \cite{Akinsanmi-etal_2020} is inclined by $i_\text{r} = 25\pm 4^\circ$ from the sky plane and rotated by $\theta = 95\pm 17^\circ$ from the transit direction. The spin-orbit obliquity $\varepsilon$ of the planet is given by $\cos\varepsilon = \cos I\cos i_\text{r} + \sin I\sin i_\text{r}\cos\theta$, where $I = 89.97\pm 0.01^\circ$ is the orbital inclination of HIP\,41378\,f.} $\varepsilon=92\pm 7^\circ$.

In order to compute the orbital precession modes of HIP\,41378\,f, our choice of prior for the masses and semi-major axes of the planets must reflect our partial knowledge of the HIP\,41378 system. We sort the planets by increasing orbital periods such that the indexes $k=(1,2,3,4,5,6)$ correspond to the planets (b, c, g, d, e, f). We assume all masses and semi-major axes to have Gaussian distributions centred on the best-fit values of \cite{Santerne-etal_2019} given in Table~\ref{tab:param}. Planet~d needs a specific treatment: even though its period has tentatively been confirmed by \cite{Grouffal-etal_2022}, it has still not been detected by the radial velocity method, so its mass is highly uncertain. We choose to remain as agnostic as possible as regards its mass, and draw it from a Gaussian fit to the mass-radius distribution of all known exoplanets having a radius between $3$ and $4$~$R_\oplus$ and a mass measurement. From the Nasa Exoplanet Archive\footnote{\url{https://exoplanetarchive.ipac.caltech.edu}} on date 2022-11-23, we obtain a central mass value of $12.7$~$M_\oplus$ and a standard deviation of $6.0$~$M_\oplus$. The high tail of this distribution may not be compatible with radial velocity measurements; yet, this broad interval gives us confidence that the actual mass of planet~d is contained in our analysis. The low tail of the distribution  (from which we cut the portion $<0.1$~$M_\oplus$) corresponds to cases in which planet~d barely exists at all. The system may also contain additional massive planets that have not been discovered yet. Hence, we stress that the analysis below represents our current knowledge of the system and it may need to be revisited in the future.

\begin{table*}
   \caption{Parameters of planets in the HIP\,41378 system used in this article.}
   \label{tab:param}
   \centering
   \begin{tabular}{ccrrrr}
      \hline
      \hline
      $k$  & name & $M_k$ ($M_\oplus$) & $P_k$ (day) & $a_k$ (au) & $I_k$ ($^\text{o}$) \\
      \hline   
        $1$ &  b  & $6.89\pm 0.88$ &  $15.57208\pm 0.00002$ & $0.1283\pm 0.0015$ & $88.75\pm 0.13$ \\
        $2$ &  c  & $4.4\pm 1.1$   &  $31.70603\pm 0.00006$ & $0.2061\pm 0.0024$ & $88.48\pm 0.07$ \\
        $3$ &  g  & $7.0\pm 1.5$   &  $62.06\pm 0.32$       & $0.3227\pm 0.0036$ & $88$ \\
        $4$ &  d  & $12.7\pm 6.0$  & $278.3618\pm 0.0005$   & $0.88\pm 0.01$     & $89.80\pm 0.02$ \\
        $5$ &  e  & $12\pm 5$      & $369\pm 10$            & $1.06\pm 0.03$     & $89.84\pm 0.07$ \\
        $6$ &  f  & $12\pm 3$      & $542.0798\pm 0.0002$   & $1.37\pm 0.02$     & $89.97\pm 0.01$ \\
      \hline
   \end{tabular}
   \tablefoot{Parameters are the mass $M_k$, period $P_k$, semi-major axis $a_k$, and inclination $I_k$ with respect to the sky plane. By convention, inclinations are given with values $I_k\leqslant 90^\circ$, but transit and radial velocity data cannot discriminate between an inclination value $I_k$ or $180^\circ-I_k$. Uncertainties are assumed to be Gaussian; quoted intervals are $1\sigma$. Parameters come from the preprint by \cite{Santerne-etal_2019} except the mass of planet~d (see text). The star mass is taken to be $1.16\pm 0.04$~$M_\odot$ for consistency with the other parameters of \cite{Santerne-etal_2019}. The inclination of the non-transiting planet~g is assumed to be $88^\circ$ as in \cite{Santerne-etal_2019}.}
\end{table*}

For the HIP\,41378 system as considered in Table~\ref{tab:param}, a look at the diagonal and off-diagonal values in the Lagrange-Laplace matrix $B$ reveals a peculiar hierarchical configuration. The system is composed of two weakly coupled subsystems: \emph{i)} the inner subsystem (planets~1-2-3) is characterised by planets~1 and 3 interacting with each other and affecting the motion of the low-mass planet~2; and \emph{ii)} the outer subsystem (planets 4-5-6) is made of the two strongly coupled planets~4 and 5, interacting as a whole with planet~6.

\begin{figure*}
   \includegraphics[width=\textwidth]{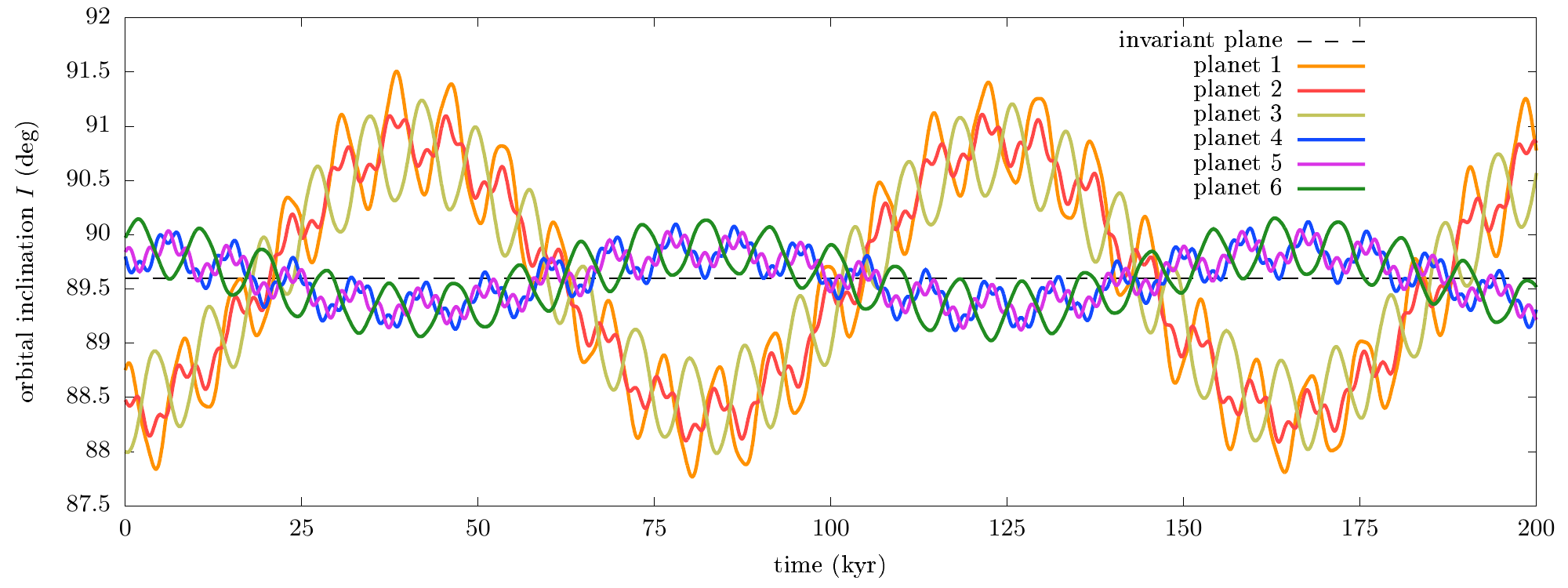}
   \caption{Example of inclination evolution of the six planets in the HIP\,41378 system. In this example, the Lagrange-Laplace equation is solved using the nominal masses and semi-major axes of all planets given in Table~\ref{tab:param}. The initial conditions $\zeta_k$ are set from the nominal inclinations $I_k$ (all assumed to be $I_k\leqslant 90^\circ$) and random longitudes of node $\Omega_k$ in a $0.2^\circ$-wide interval; this choice is commented in Sect.~\ref{sec:capture}.}
   \label{fig:exprec}
\end{figure*}

This peculiar hierarchy can be visualised by solving the Lagrange-Laplace system a first time using reasonable values for the parameters. The exact values of the parameters do not matter for now; this first step only serves as a guide to identify the frequencies and choose an adequate naming convention. Figure~\ref{fig:exprec} shows an example obtained from the nominal masses and semi-major axes of the planets. We name the proper frequencies according to their qualitative role in the dynamics: $s_1$ is the precession frequency of planets~1 and 3 about their total angular momentum vector; $s_2$ is the precession frequency of the low-mass planet~2 under the action of planets~1 and 3; $s_3$ is the slow rigid precession of the inner and outer subsystems (planets 1-2-3 and 4-5-6); $s_4$ is the precession frequency of planets~4 and 5 about their total angular momentum vector; $s_5$ is identically zero; $s_6$ is the precession frequency of planet~6 and planets~4-5 about their total angular momentum vector. We stress that all precession modes actually appear in the dynamics of all planets (see Eq.~\ref{eq:LLsol}), but this qualitative description gives us a good idea of the relative importance of each term in the orbital evolution of each planet.

In order to compute the probability density function of each frequency $s_j$ given our current knowledge of the planetary system, we drew $10^6$ realisations of the star's mass and planets' masses and semi-major axes. For each of these realisations, we computed the eigenvalues of the Lagrange-Laplace matrix $B$ and identified them to the frequencies $s_j$ according to their qualitative role described above. In practice, this identification can be made by choosing fictitious initial conditions $\zeta_k(t=0)$ designed to magnify the specific term we are looking for. For instance, the frequency $s_3$ would appear as strongly dominant for all planets if we set $\zeta_k(t=0)=0$ for $k=\{1,2,3\}$ and $\zeta_k(t=0)=\sqrt{2}/2$ for $k=\{4,5,6\}$. Then, one may identify $s_2$ as the dominant term in the solution of planet~2 by setting $\zeta_2(t=0)=\sqrt{2}/2$ and $\zeta_k(t=0)=0$ for $k\neq 2$, etc. This way, all frequencies can be correctly identified one by one. Moreover, we remind the reader that the frequencies $s_j$ only depend on the masses and semi-major axes of the planets, so they do not depend on the fictitious initial conditions chosen here, and they are not plagued with our ignorance of the actual orientations of the planets' orbital planes.

\begin{figure}
   \includegraphics[width=\columnwidth]{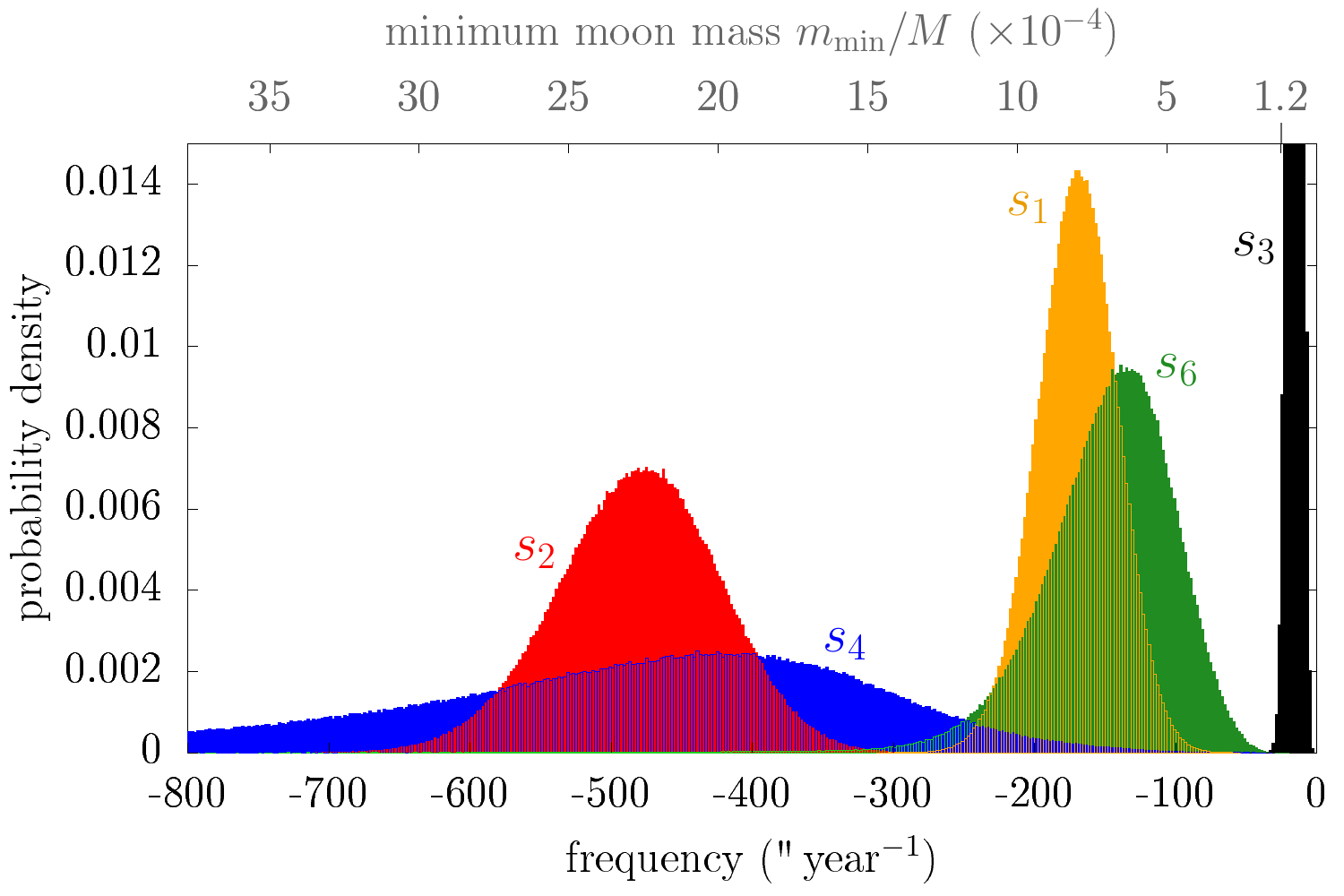}
   \caption{Probability density of the inclination proper modes of the HIP\,41378 system. Histograms are built from $10^6$ realisations of the Lagrange-Laplace system with the mass and semi-major axis uncertainties in Table~\ref{tab:param}. The histogram for frequency $s_4$ has a long tail extending beyond the left border of the figure (with $99.7\%$ occurrences above $-1850$~$''$\,year$^{-1}$ and $95.4\%$ above $-1000$~$''$\,year$^{-1}$). The histogram for frequency $s_3$ peaks above the top border of the figure. Frequency $s_5$ is identically equal to zero from the conservation of angular momentum. The upper axis shows the minimum moon mass needed for HIP\,41378\,f to be fully tilted through a resonance with a given frequency value (see Sect.~\ref{sec:moon}).}
   \label{fig:freq}
\end{figure}

Figure~\ref{fig:freq} shows the frequency distribution for each inclination proper mode obtained from our $10^6$ realisations of the system. Frequency $s_4$ has a broad distribution due to the large uncertainties in the masses of planets~d and e. Frequency $s_3$, on the contrary, is very peaked, which means that the hierarchy of the two subsystems is a robust property of the HIP\,41378 system -- unless it contains additional massive planets yet to be discovered. In order to quantify the relative importance of each parameter in the value of each frequency, a correlation analysis can be performed on our large sample of realisations. Here, the small spread in frequency $s_3$ results to be essentially due to the uncertainty in the mass of planet~d (see Appendix~\ref{asec:correl}).

As illustrated in Fig.~\ref{fig:exprec}, frequency $s_3$ is expected to have a strong contribution in the motion of all planets. Next to it, the dominant inclination proper mode of planet~f has frequency $s_6$. This frequency would produce a strong (if not the strongest) secular spin-orbit resonance for this planet. Figure~\ref{fig:freq} shows that despite observational uncertainties, frequency $s_6$ has a relatively peaked distribution. Its most probable value is $-136$~$''$\,year$^{-1}$, with $68.3\%$ occurrences within $[-181,-97]$~$''$\,year$^{-1}$, $95.4\%$ occurrences within $[-241,-65]$~$''$\,year$^{-1}$, and $99.7\%$ occurrences within $[-405,-35]$~$''$\,year$^{-1}$. As shown in Appendix~\ref{asec:correl}, the value of $s_6$ is essentially set by the mass of the perturbing planet~e, with a Spearman correlation coefficient $\rho_\mathrm{S}\approx -0.8$. The value of $s_6$ is only weakly ($|\rho_\mathrm{S}|\lesssim 0.3$) correlated with the parameters of planet~f itself. This low correlation allows us to investigate different values for the frequency $s_6$ independently of the mass and semi-major axis of planet~f (that we fix, from now on and in the rest of the article, to their nominal values in Table~\ref{tab:param}).

\section{Properties of the hypothetical former moon}\label{sec:moon}

Knowing the dominant harmonics in the orbital precession of a planet, Eq.~\eqref{eq:condtilt} gives the conditions required to tilt the planet and form a ring through the tidal migration and disruption of a moon. In addition to the mass and orbital elements of the planet, Eq.~\eqref{eq:condtilt} depends on the planet's normalising radius $R$, its oblateness coefficient $J_2$, and the product $\omega\lambda$. For a given super-puff exoplanet, we may assume that the anomalous planet's density is entirely due to the existence of a ring; therefore, the value of $R$ can be chosen so as to produce a conventional bulk density (e.g. that of Uranus or Neptune). In the specific case of HIP\,41378\,f, \cite{Akinsanmi-etal_2020} show that, under the ring hypothesis, its true radius would be $3.7_{-0.2}^{+0.3}$~$R_\oplus$. Hence, we adopt the value $R=3.7$~$R_\oplus$ below as our normalising radius.

For given values of the parameters $J_2$ and $\omega\lambda$, Eq.~\eqref{eq:condtilt} provides a direct relation between the frequency $\nu_j$ of the resonance and the minimum mass $m_\mathrm{min}$ of the former moon. Even though $J_2$ and $\omega\lambda$ are completely unknown for exoplanets, we know that they are related, and in first approximation $J_2\propto\omega^2$ (planets spinning faster are more flattened; see e.g. \citealp{Chandrasekhar_1969}). For a given moon mass $m$, the condition $|\nu_j|\leqslant p\eta/2$ in Eq.~\eqref{eq:condtilt} corresponds to a power law $J_2\propto\omega^{5/2}$. Because of the coincidental near match between these two exponents ($2$ and $5/2$), our total ignorance of $J_2$ and $\omega\lambda$ does not affect much our estimate of $m_\mathrm{min}$: we may just set $J_2$ and $\omega\lambda$ to realistic values (e.g. obtained from the Solar System planets) and be assured to obtain relevant results -- unless the planet has a particularly exotic internal structure which violates $J_2\propto\omega^2$. This property is verified in Appendix~\ref{asec:param} in the case of planet HIP\,41378\,f. As the mass and radius proposed by \cite{Akinsanmi-etal_2020} for HIP\,41378\,f are relatively close to those of Uranus, we choose to apply Eq.~\eqref{eq:condtilt} using the parameters $J_2$ and $\omega\lambda$ of Uranus (see e.g. \citealp{Yoder_1995}).

Independently of the resonance considered, Eq.~\eqref{eq:condtilt} can be fulfilled only if the mass parameter $\eta$ of the moon is $\eta\geqslant 2$. Using the $J_2$ value of Uranus, this condition translates into $m/M\geqslant 1.2\times 10^{-4}$. This is the minimum mass ever that the former moon of HIP\,41378\,f should have had. For a larger moon, the minimum mass $m_\mathrm{min}$ needed to tilt the planet is proportional to the frequency $\nu_j$ of the considered resonance. The top horizontal axis in Fig.~\ref{fig:freq} shows the values of $m_\mathrm{min}$ computed from Eq.~\eqref{eq:condtilt} using the parameters $J_2$ and $\lambda\omega$ of Uranus (the tics start at $1.2\times 10^{-4}$ and go from right to left).

The characteristic spin-axis precession rate of HIP\,41378\,f computed from Eq.~\eqref{eq:p} is $p\approx 25''$\,year$^{-1}$. According to the left inequality in Eq.~\eqref{eq:condtilt}, this value almost certainly rules out a resonance with frequency $s_3$, because frequency $s_3$ sharply peaks at $s_3=-15.7''$\,year$^{-1}$ (see Fig.~\ref{fig:freq}). The fact that $p>|s_3|$ means that the $s_3$ resonance is located in the green portion of Fig.~\ref{fig:precrate}; therefore no capture from a low obliquity is possible in this resonance whatever the mass of the moon. Frequency $s_6$, on the contrary, is the closest resonance reachable by HIP\,41378\,f. This resonance is expected to be strong for planet~f, if not the strongest (see Sect.~\ref{sec:orbit}). Figure~\ref{fig:freq} shows that a capture and full tilting within the $s_6$ resonance requires a moon with minimum mass ratio ranging between about $2\times 10^{-4}$ and $10\times 10^{-4}$. This corresponds to an absolute mass ranging roughly between Triton's mass and the mass of our Moon, respectively. More precisely, when the parameters $J_2$ and $\omega\lambda$ of Uranus are assumed for HIP\,41378\,f, the value of frequency $s_6=-136^{+101}_{-269}$~$''$\,year$^{-1}$ obtained in Sect.~\ref{sec:orbit} translates into a minimum moon mass $m_\mathrm{min}/M=6^{+13}_{-5}\times 10^{-4}$ ($3\sigma$ uncertainty).

This mass range seems realistic when viewed in the context of the regular moons of the Solar System giant planets. For comparison, the moon-to-planet mass ratio of Titan is $2\times 10^{-4}$, and the summed masses of the largest moons of Jupiter and Uranus yield ratios of about $2\times 10^{-4}$ and $1\times 10^{-4}$, respectively. This similarity among planets motivated the work of \cite{Canup-Ward_2006}, who found that the formation mechanism of moons around the Solar System giant planets may naturally lead to a common mass scaling, with final mass ratios of a few times $10^{-4}$. Yet, these results do not rule out the existence of larger moons, either because of differing external conditions during their formation, or because of different formation processes (see e.g. the discussion by \citealp{Saillenfest-etal_2022}).

In order to fully incline the planet starting from a low obliquity, the distance that the migrating moon needs to cover depends on the resonance considered, but Fig.~\ref{fig:precrate} shows that one can expect in general a migration from $a_\mathrm{m}\approx 0.5\,r_\mathrm{M}$ to $1\,r_\mathrm{M}$. Using the $J_2$ value of Uranus, Eq.~\eqref{eq:rM} gives a characteristic length $r_\mathrm{M}\approx 11$~$R$ for planet HIP\,41378\,f, which implies that the moon would need to migrate from roughly $5$ to $10$~$R$. Given that $r_\mathrm{M}$ is proportional to $J_2^{1/5}$, other realistic values of $J_2$ may change these distances by a small amount (see discussion in Appendix~\ref{asec:param}).

The HIP\,41378 system is $2.1^{+0.4}_{-0.3}$~Gyr-old \citep{Lund-etal_2019}. As the whole tilting mechanism must have been completed before today, the required migration range for the moon can be translated into a minimum migration rate. In the case of HIP\,41378\,f, we obtain a migration rate of about $6$~cm\,year$^{-1}$ in average. This velocity is comparable to the Moon's migration rate from the Earth \citep{Williams-Boggs_2016}, and about two times less than the migration rates of Ganymede from Jupiter \citep{Lainey-etal_2009} or Titan from Saturn \citep{Lainey-etal_2020}. In order to power this migration through tidal dissipation within the planet, classical formulas with constant parameters (see e.g. \citealp{Efroimsky-Lainey_2007}) imply that the planet's dissipation coefficient needs to be higher than $k_2/Q\approx 3\times 10^{-5}$ for a moon mass $m/M=2\times 10^{-4}$, and higher than $k_2/Q\approx 6\times 10^{-6}$ for a moon mass $m/M=10^{-3}$. For comparison, the value measured for Jupiter's satellite Io is $k_2/Q=1.102\pm 0.203\times 10^{-5}$ \citep{Lainey-etal_2009}, and the value measured globally for Saturn's main satellites is $k_2/Q=1.59\pm 0.74\times 10^{-4}$ \citep{Lainey-etal_2017} with a large spread for individual moons extending to much higher values (see \citealp{Lainey-etal_2020,Jacobson_2022}).

\section{Adiabatic resonance capture}\label{sec:capture}

The analysis above shows that when assuming realistic values for the unknown parameters $J_2$ and $\omega\lambda$, the constraints obtained for the planet HIP\,41378\,f and its hypothetical former moon match well the properties expected for giant planets and moons (i.e. distance, mass, migration rate, and tidal dissipation), at least when viewed in the context of the Solar System. Yet, in order for a planet to be captured and adiabatically tilted within a given resonance, this resonance must be large enough. The width of secular spin-orbit resonances scales as the square root of the amplitude of the term in the orbital series (see Eq.~\ref{eq:zeta}). The $s_6$ term is expected to be among the dominant terms for planet HIP\,41378\,f, but its amplitude may still be small, depending on the mutual inclinations between the planets' orbital planes. In order to compute the mutual inclinations of the planets, we need their orbital inclinations $I_k$ and longitudes of ascending nodes $\Omega_k$.

As shown in Table~\ref{tab:param}, the orbital inclinations $I_k$ of transiting planets with respect to the sky plane are tightly constrained from observations, apart from the mirror degeneracy with respect to $90^\circ$. As for the longitudes of nodes $\Omega_k$ in the sky plane, they are not constrained from transit photometry, but we know that their values are likely to be close to each other. Indeed, for a given set of orbital inclinations $I_k$, mutual inclinations between the planets' orbital planes are minimum if their longitudes of node $\Omega_k$ are equal. As a general rule, low mutual inclinations minimise the planets' orbital excitation, and a low orbital excitation is expected in multi-planetary systems for stability reasons.

In systems observed by the transit method, low mutual inclinations are expected also because they maximise the probability of observing several transiting planets. Gravitational interactions produce a precession of the planets' orbital planes, possibly making some of them evolve in and out of transit configuration (see e.g. \citealp{Becker-Adams_2016}). Using the Lagrange-Laplace theory, it is straightforward to compute the fraction of time that a planet spends in and out of transit configuration (see e.g. Fig.~\ref{fig:exprec}). In the HIP\,41378 system as described in Table~\ref{tab:param}, only the innermost planet may possibly transit $100\%$ of the time, even if we set all the $\Omega_k$ values of the planets to be equal. Due to orbital precession, the probability to observe five transiting planets (as today) is $30\%$ at best, and the probability to observe six is lower than $5\%$. As such, the HIP\,41378 system would not be classified as `continually mutually transiting' \citep{Becker-Adams_2016,Becker-Adams_2017}.

The level of orbital excitation of a planetary system can be quantified as a function of the dispersion of their longitudes of ascending node $\Omega_k$ in the sky plane. As shown in Appendix~\ref{asec:amplitudes}, allowing for just a few degrees dispersion in $\Omega_k$ can increase the amplitude $S_j$ of several modes in Eq.~\eqref{eq:LLsol} by orders of magnitude, drastically reducing transit probabilities. In the HIP\,41378 system, the level of dispersion of the planets' longitudes of node $\Omega_k$ is therefore likely to be very small, perhaps less than $1^\circ$, but their actual values are unknown.

Here, we are interested in the possibility for a planet to be captured in secular spin-orbit resonance from a low initial obliquity. In this context, the larger the resonance, the easier the capture (see e.g. \citealp{Saillenfest-etal_2020}); hence, we actually just need a lower bound for the resonance widths, that is, a lower bound for the amplitudes $S_j$ in Eq.~\eqref{eq:LLsol}. If we show that the resonance capture operates flawlessly for this lower bound, then we can be assured that it will operate as well or even better for the true amplitudes $S_j$. To this aim, we consider that: \emph{i)} the orbital inclinations of all planets with respect to the sky plane lie on the same side of $90^\circ$, and \emph{ii)} all planets have exactly the same longitude of ascending node $\Omega_k$ in the sky plane. When applied to the HIP\,41378 system, this idealised system gives the solution shown in Table~\ref{tab:zeta} for planet~f.

\begin{table}
   \caption{Solution for the long-term inclination dynamics of planet HIP\,41378\,f given by the Lagrange-Laplace system.}
   \label{tab:zeta}
   \centering
   \begin{tabular}{rcrrr}
      \hline
      \hline
      $j$ & identification & $\nu_j$ ($''\,\text{yr}^{-1}$) & $S_j\times 10^7$ & $\phi_j^{(0)}$ ($^\text{o}$) \\
      \hline   
          $1$ & $s_5$ &    $0.000$ & $7046055$ &   $0.0$ \\
          $2$ & $s_3$ &  $-15.600$ &   $18545$ &   $0.0$ \\
          $3$ & $s_6$ & $-144.623$ &    $5682$ &   $0.0$ \\
          $4$ & $s_1$ & $-170.310$ &     $972$ & $180.0$ \\
          $5$ & $s_4$ & $-477.109$ &      $26$ & $180.0$ \\
          $6$ & $s_2$ & $-477.679$ &       $5$ & $180.0$ \\
      \hline
   \end{tabular}
   \tablefoot{All planets have their nominal masses, semi-major axes, and inclinations given in Table~\ref{tab:param}. The amplitudes $S_j$ are minimised by assuming that all planets have orbital inclinations lying on the same side of $90^\circ$ (chosen to be $I_k\leqslant 90^\circ$), and all planets have the same longitude of ascending node in the sky plane (chosen to be $\Omega_k=0$).}
\end{table}

In order to produce a resonance capture, the migration of the moon must be slow compared to the oscillations of the resonance angle, so that the parameter change is close to the adiabatic regime (see e.g. \citealp{Su-Lai_2020}). For a given resonance, the oscillation frequency near the resonance centre can be computed through Eq.~\eqref{eq:Tlib}; the frequency scales as the square root of the amplitude $S_j$. When applying Eq.~\eqref{eq:Tlib} to HIP\,41378\,f by considering the orbital series in Table~\ref{tab:zeta}, one finds that the libration period of the $s_6$ resonance angle when the separatrix appears is $T_\mathrm{lib}\approx 547\,000$~years. This value is much smaller than the age of the system ($2.1^{+0.4}_{-0.3}$~Gyr; see \citealp{Lund-etal_2019}). Therefore, even when considering the minimum possible width of the resonance, the available time span is more than enough for the planet to oscillate many times within the $s_6$ resonance, allowing an adiabatic drift to occur within this resonance.

\begin{figure*}
   \includegraphics[width=\textwidth]{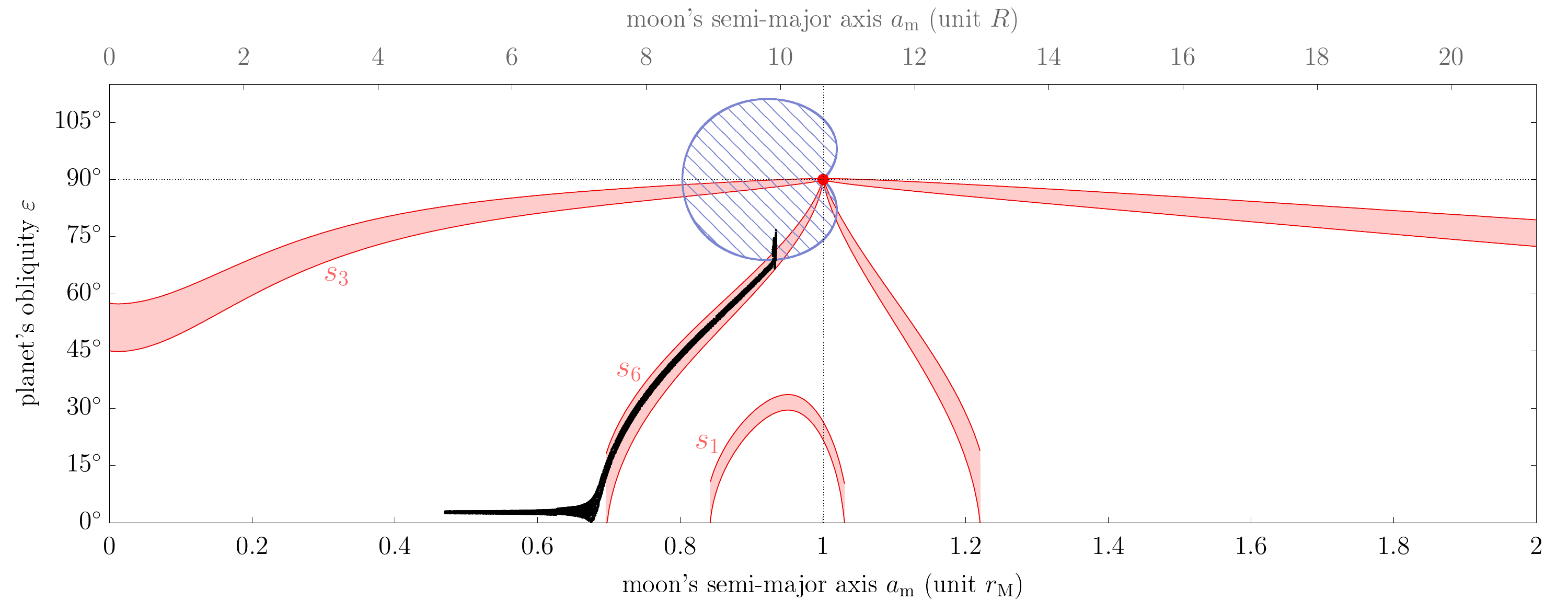}
   \caption{Example of tidal evolution of the planet HIP\,41378\,f and a hypothetical former moon. The mass of the moon is chosen to be $m/M=7\times 10^{-4}$. The moon migrates away at constant rate emulating a tidal parameter $k_2/Q=10^{-5}$. The trajectory of the system is shown in black; it goes from the leftmost to the rightmost point in about $1.3$~Gyr. The available resonances are shown in pink, with their separatrices in red; they are labelled with the frequencies $s_k$ of the corresponding modes (see Sect.~\ref{sec:orbit}). In this example, the resonances have the minimum possible widths according to the planets' orbital elements in Table~\ref{tab:param}. In the hatched blue region, the moon is unstable (same as Fig.~\ref{fig:precrate}). The top axis shows the moon distance in unit of the planetary radius.}
   \label{fig:extilt}
\end{figure*}

This point can be verified by performing a numerical integration of the coupled equations of motion of the planet's spin axis and the orbit of its moon. We used the same setting as \cite{Saillenfest-etal_2022}: we integrated the secular equations of \cite{Correia-etal_2011} expanded at quadrupole order, and forced the orbital evolution of the planet with the quasi-periodic series in Table~\ref{tab:zeta}. A typical example of evolution is displayed in Fig.~\ref{fig:extilt}. In this example, the mass of the moon is $m/M=7\times 10^{-4}$ (i.e. about the mass of Jupiter's moon Europa), and we made the moon migrate outwards at a constant rate, chosen to emulate a tidal parameter $k_2/Q\approx 10^{-5}$. For such a tidal parameter, the moon is expected to migrate from a distance $a_\mathrm{m}=5$~$R$ to a distance $a_\mathrm{m}=10$~$R$ in about $1.2$~Gyr. The planet was initialised with an obliquity of $0.05$~rad and a random precession phase. The eccentricity of the moon and its inclination with respect to its local Laplace plane were both initialised to $10^{-4}$, with random argument of pericentre and longitude of ascending node. As expected, Fig.~\ref{fig:extilt} shows that the adiabatic capture and tilting in resonance $s_6$ is guaranteed on a gigayear timescale. Due to the large separation between timescales, the obliquity oscillations of the planet inside the resonance are not even noticeable in the figure, but they build up in the curve width.

When the system reaches the unstable region, the eccentricity of the moon increases rapidly, which produces chaotic jumps in the planet's obliquity. Indeed, near the border of the unstable region, the timescale for the moon's eccentricity to be multiplied by $100$ is a few times the characteristic timescale $\tau$ defined in Eq.~\eqref{eq:kappa}. Here, one obtains $\tau\approx 100$~yr, which means that the eccentricity increase is extremely fast compared to the planet's spin-axis precession timescale ($T\approx 52\,000$~yr), to the oscillations of the planet inside the resonance ($T_\mathrm{lib}\approx 547\,000$~yr), or to the tidal eccentricity damping of the moon (whose timescale is a few millions of years; see e.g. \citealp{Murray-Dermott_1999}).

The simulation in Fig.~\ref{fig:extilt} is stopped when the moon's pericentre goes below the Roche limit of the planet. At this point, the moon is expected to be disrupted into pieces which would rapidly reorganise into an equatorial disc confined inside the Roche limit (see e.g. \citealp{Canup_2010,Hyodo-etal_2017}). As the moon is lost, the planet is suddenly released from any kind of spin-orbit coupling, and its obliquity remains permanently frozen. In the example shown in Fig.~\ref{fig:extilt}, the final obliquity of the planet is about $77^\circ$. This value is roughly compatible at $2\sigma$ with the obliquity $\varepsilon=92\pm 7^\circ$ proposed by \cite{Akinsanmi-etal_2020}. However, we stress that the final obliquity of the planet is the result of a chaotic phase; its value strongly depends on initial conditions, on the mass of the moon, and on the widths of nearby secular spin-orbit resonances \citep{Saillenfest-etal_2022}. More massive moons and larger resonances increase the obliquity excitation of the planet during the chaotic phase. Due to chaos, obliquity values larger than $90^\circ$ can be reached, but the detailed exploration of possible outcomes would require a precise knowledge of the orbital dynamics of the planet. Without this knowledge, we can only conclude that the obliquity of the planet ends up within the hatched blue region in Fig.~\ref{fig:extilt}, that is, between about\footnote{The closed-form expression for the border of the unstable region is $\cos^2\varepsilon = (51 + 25\sqrt{3})/726$; see \cite{Saillenfest-Lari_2021}.} $70^\circ$ and $110^\circ$.

\section{Discussions}\label{sec:discussion}

\subsection{Refining the tilting mechanism}
Under the ring hypothesis, we have presented a proof of concept for producing the unusual configuration proposed for super-puff exoplanets through the tidal migration of a former moon. We have considered the effect of a single massive moon on the planet's spin axis dynamics. This does not mean that the planet only had one moon -- we expect it to possibly have many -- but that this big moon gathered most of the mass of the satellite system, similarly to Titan around Saturn. Now that this big moon is lost, the remaining moons (either pre-existing or formed in the debris ring) are expected to be very small and undetectable with current facilities.

The presence of several pre-existing big moons, as the Galilean satellites around Jupiter, would complicate the picture outlined here. Through their mutual gravitational perturbations, several massive moons could either inhibit or facilitate the tilting process (see \citealp{Saillenfest-etal_2022}). The exploration of this more complicate scenario is out of the scope of this article. More generally, additional work can refine the scenario proposed here for a given target exoplanet, including the efficiency of ring formation, the distribution of possible final obliquities, and the combined effect of several massive moons. However, this level of detail would require an in-depth knowledge of the orbital dynamics of the planetary system.

In the case of HIP\,41378\,f, confirmed periods and masses are still missing for planets~d, planet~e, and the candidate planet~g. The analysis presented here reflects our current understanding of the system, and some results may change in case of substantial modifications in the system's hierarchy. Our correlation analysis shows that planets~d and g are only weakly coupled with the frequency $s_6$ of the resonance involved. A mass measurement for these planets would therefore not alter much the picture outlined above. However, substantial changes could be produced if future observations reveal a substantially different mass or period for planet~e, or if the system contains an additional outer planet; the calculations presented here should therefore be updated. In this respect, the simplicity of the analytical formulas involved is a great advantage.

\subsection{The true nature of super-puffs}
Future characterisation of super-puff exoplanets is fundamental to assess the actual nature of their anomalously large radii. Unfortunately, due to the nearly face-on configuration of the proposed ring, an unambiguous detection of the ring by transit photometry or by the Rossiter-McLaughlin effect would be challenging with current instruments \citep{Akinsanmi-etal_2020}. Spectroscopic observations are much more promising. Even though the spectra of several super-puffs have been revealed to be featureless in near infrared \citep{LibbyRoberts-etal_2020,Chachan-etal_2020,Alam-etal_2022}, rings are expected to be transparent in far infrared, which would strongly reduce the transit depth of the planet. As noted by \cite{Alam-etal_2022}, mid-infrared observations by the \emph{JWST} would be enough to break the degeneracy between high-altitude hazes, a high-metallicity atmosphere, or the ring hypothesis. The nominal \emph{JWST} mission offers only two opportunities to observe a transit of HIP\,41378\,f: October 2025 and March 2027. Considering their high scientific value, these opportunities should not be missed. In addition, the high cadence and high photometric resolution of the future \emph{PLATO} mission may allow small distortions in the transit light curve to be detected (due to the non-zero inclination of the ring with respect to the sky plane and/or to a possible thin inner gap in the ring; see \citealp{Akinsanmi-etal_2020}).

\subsection{The rarity of enlarged planets}
Due to the generic nature of the mechanism presented here, one may wonder why in this case we do not observe many distant exoplanets with anomalously large radii. This rarity can be explained by several factors. First, the transit and radial-velocity methods are strongly biased towards the detection of short-period exoplanets \citep{Perryman_2018}. In this regard, the detection of HIP\,41378\,f with a period of $542$~days is already an exception (the transit probability is $0.5\%$). In turn, the long-period planets observed in direct imaging are strongly biased towards young systems, which cannot have gone through the gigayear adiabatic tilting process described here. As of today, this leaves us with only a handful of exoplanet detections for which this mechanism may have played a role.

\begin{figure}
   \includegraphics[width=\columnwidth]{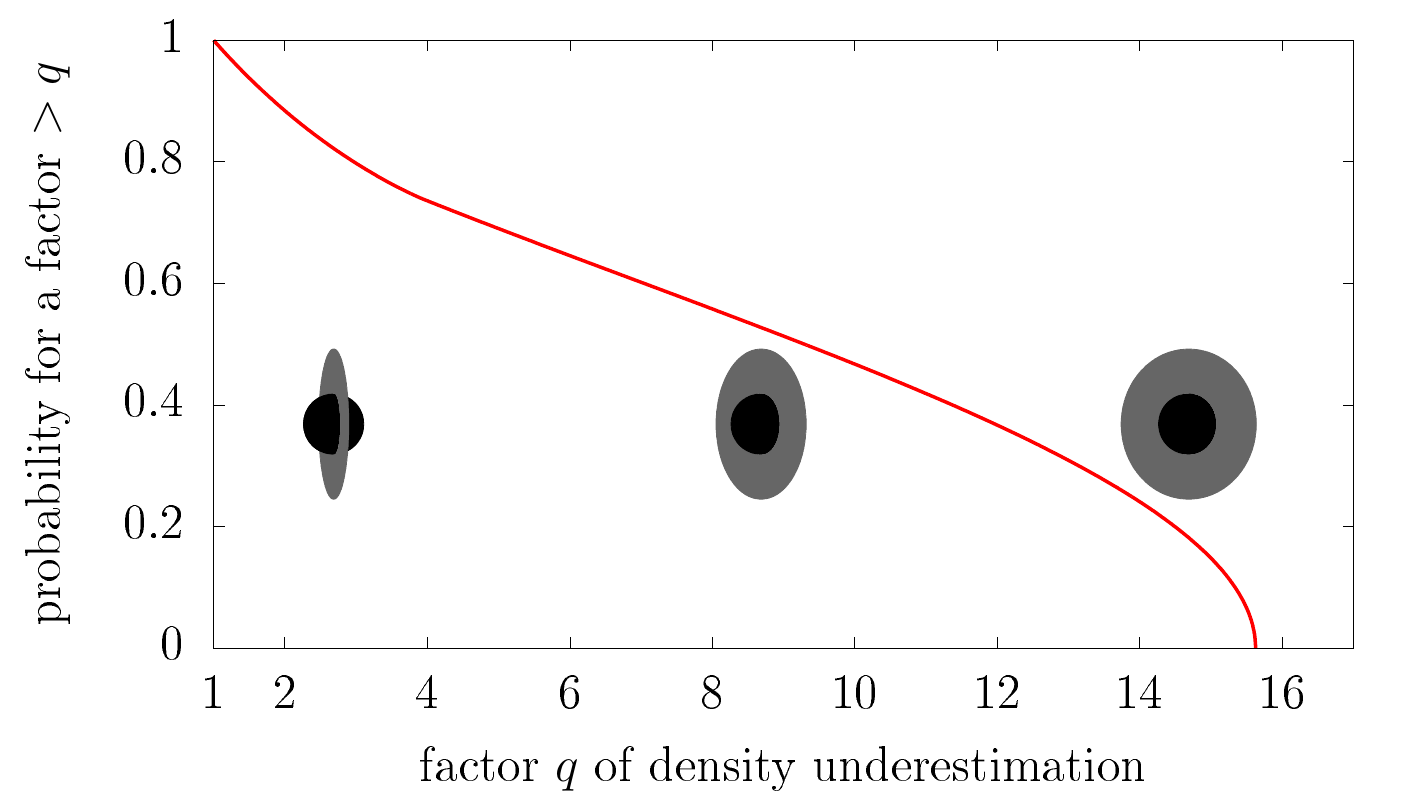}
   \caption{Probability of underestimating the density of a transiting exoplanet by a factor $q$ due to the presence of an opaque ring with outer radius $2.5$~$R$ and no inner gap. The enhanced transit depth due to the ring is supposed to be fully misinterpreted as an enlarged planetary radius. The red curve is obtained by computing the ring inclination required to divide the measured planet density by a factor $q$, and by assuming that the inclination of the ring (or equivalently, its precession angle $\psi$; see text) is uniformly distributed between $0$ and $2\pi$. The three planet pictures show the approximate geometries corresponding to the factor $q$ in abscissa. The probability goes from $1$ at $q=1$ (exact edge-on configuration) to $0$ at $q=2.5^3$ (exact face-on configuration). In case it possesses a ring, planet HIP\,41378\,f would have $q\approx 13$ \citep{Akinsanmi-etal_2020}.}
   \label{fig:density}
\end{figure}

The second rarity factor is geometric: a strong radius enhancement able to cast suspicion requires a roughly face-on ring. The mechanism proposed here produces a final planetary obliquity more or less equal to $90^\circ$, which is a necessary condition for observing a transiting face-on ring, but is not sufficient: the precession phase $\psi$ of the planet must also have an adequate value. For a ring with typical radius $2.5$~$R$, the increase in transit depth leading to underestimating the planet density by a factor $q>10$ requires a precession phase within $\pm 40^\circ$ of the exact face-on configuration (see e.g. \citealp{Zuluaga-etal_2015}). As shown in Fig.~\ref{fig:density}, this occurs about $45\%$ of the time. This fraction is lowered if we consider the ring to have an inner optically thin gap similar to Saturn's ring.

Finally, even though the mechanism described in this article is generic, not all giant planets are expected to reach the final instability phase in only a few gigayears. Depending on the initial configuration of their moons and the geometry of the available resonances, the planet's obliquity may only have time to increase by a few tens of degrees during its lifetime. In the Solar System, which is aged $4.5$~Gyr, only Uranus may have completed the final stage today \citep{Saillenfest-etal_2022}. In contrast, Jupiter is only starting the tilting phase\footnote{As Jupiter possesses four massive moons interacting with each other, its tilting process is somewhat different from what is presented here. Jupiter may never be able to reach an obliquity close to $90^\circ$ even if it was given infinite time.} \citep{Saillenfest-etal_2020}, while Saturn is seemingly halfway in \citep{Saillenfest-etal_2021a,Saillenfest-etal_2021b} -- and it may have recently been ejected from resonance (see \citealp{Wisdom-etal_2022}).

Hence, even though many exoplanets are probably affected by this mechanism, the conjunction of observational biases, ring geometry, and the long timescales at play drastically reduces the probability of detecting targets as exquisite as HIP\,41378\,f. In this regard, the future \emph{PLATO} mission is particularly promising, as its observing strategy is tailored to long-period planets, and it will be accompanied by an intensive radial-velocity follow-up to get accurate planet masses and detect possible non-transiting companions. Hopefully, the \emph{PLATO} discoveries will enable us to estimate the fraction of the exoplanet population that may have gone through the mechanism described in this article.

\section{Conclusion}\label{sec:conclusion}
The apparent enlarged radius of some long-period exoplanets may be due to the presence of a ring observed roughly face on \citep{Piro-Vissapragada_2020,Akinsanmi-etal_2020}. Despite their unconventional configuration, such hypothetical rings and the nearly $90^\circ$ obliquity of their host planets can be the natural end state of former migrating moons. This mechanism involves the capture of the planet in secular spin-orbit resonance as the moon migrates away on a gigayear timescale. The planet is then gradually tilted until the moon is destabilised and may be destructed into a debris disc.

For a given exoplanet, the plausibility of this formation mechanism can be assessed through simple analytical calculations. First, we need to determine the list of secular spin-orbit resonances that may tilt the planet. The frequencies $\nu_j$ of the main orbital precession harmonics of the planet can be obtained through the Lagrange-Lagrange theory; in this theory, orbital frequencies are the eigenvalues of a matrix which depends only on the masses and spacings of the planets contained in the system. The probability density function of each frequency can be built from numerous realisations of the system (e.g. $10^6$ or more) which are sampled according to our uncertainties on the parameters. Simple correlation analysis can then quantify the influence of each planet in the frequency values.

Then, for each frequency $\nu_j$, the simple formula in Eq.~\eqref{eq:condtilt} gives the minimum mass of a moon that the planet must have in order to trigger an adequate secular spin-orbit resonance. This formula depends on the unknown parameters $J_2$ and $\omega\lambda$ of the planet, but thanks to the approximate relation $J_2\propto\omega^2$, this lack of knowledge only weakly affects the final result. The moon-to-planet mass ratio obtained is the first plausibility check of this dynamical mechanism. Moons with mass ratio $m/M\sim 10^{-4}$ or smaller are expected to be ubiquitous around gaseous planets (see e.g. \citealp{Canup-Ward_2006}). Substantially larger moons cannot be categorically ruled out, but they would require non-generic formation pathways such as captures or giant impacts, and are therefore much less likely (see e.g. \citealp{Kipping_2014}).

A second consistency check is provided by the age of the planetary system considered. The Laplace radius of the planet (see Eq.~\ref{eq:rM}) sets the distance over which the moon needs to migrate to fully tilt the planet. The migration range obtained must have been covered by the moon in a smaller timespan than the age of the system. As the migration of moons is powered by tidal dissipation inside the planet, the required distance and migration timescale can be translated into a tidal parameter $k_2/Q$ for the planet. Expected values are of the order of $10^{-5}$ to $10^{-4}$ from a Solar System perspective \citep{Lainey-etal_2009,Lainey-etal_2017}.

The last plausibility check is the consistency of timescales between the age of the planetary system and the hypothesis of an adiabatic capture into resonance. An adiabatic capture requires the libration period inside the resonance to be much shorter that the age of the system, such that many oscillations of the resonance angle may possibly have occurred during the tilting of the planet. The characteristic libration frequency is given in Eq.~\eqref{eq:Tlib}; it depends on the frequencies $\nu_j$ obtained above, but also on the amplitudes $S_j$ of the corresponding harmonics in the planet's orbital precession spectrum. Using the Lagrange-Laplace theory, the computation of these amplitudes requires to know the inclinations $I_k$ and longitudes of nodes $\Omega_k$ of the planets (e.g. measured in the sky plane). As the libration frequency scales as the square root of the amplitude of the resonant term, only a lower bound for $S_j$ is actually needed. This lower bound can be obtained even in case the longitudes of nodes $\Omega_k$ of the planets are unknown, allowing one to compute a maximum value for the libration period inside the resonance considered.

We applied this methodology to the planet HIP\,41378\,f, and obtained that all consistency checks are fulfilled. In order to tilt the planet through an adequate resonance, the hypothetical exomoon must have had a moon-to-planet mass ratio ranging from $m/M\gtrsim 2\times 10^{-4}$ to $10\times 10^{-4}$, that is, a mass comparable to that of Neptune's moon Triton, Jupiter's moon Europa, or to that of our own Moon. Even though such small exomoons are very hard to detect due to the weakness of their observational signals \citep{Kipping_2014}, we expect them to be ubiquitous around giant exoplanets. Provided that the exomoon was initially formed at a distance of about $3$ to $10$ planetary radii (similarly to Jupiter's moons Io or Europa), its outward migration leads to a guaranteed capture of HIP\,41378\,f in a secular spin-orbit resonance. The migration timescale required for the moon is found to be in line with what is observed in the Solar System, with a corresponding tidal dissipation factor $k_2/Q$ larger than $2\times 10^{-5}$ (for the smallest possible moon) or larger than about $6\times 10^{-6}$ (for a bigger moon). Finally, the libration timescale inside the resonance is found to be orders of magnitudes smaller than the age of the system ($2.1^{+0.4}_{-0.3}$~Gyr; see \citealp{Lund-etal_2019}), allowing for the whole tilting mechanism to have possibly occurred.

All these requirements are confirmed by an example of fully coupled numerical integration of the planet's spin axis and the moon's orbit. The planet's spin axis is gradually tilted until its obliquity $\varepsilon$ reaches values in the interval $[70^\circ,110^\circ]$, and its moon becomes unstable \citep{Tremaine-etal_2009,Saillenfest-Lari_2021}. Due to the short instability timescale of the exomoon ($\tau\approx 100$~yr in the case of HIP\,41378\,f), its eccentricity increase is likely to cause catastrophic events, such as collision chains between small inner moons or a tidal disruption of the moon itself when its pericentre goes below the planet's Roche limit (see e.g. \citealp{Canup_2010,Hyodo-etal_2017,Wisdom-etal_2022}). Hence, we argue the dynamical mechanism described here, which may be responsible for the tilting of planet HIP\,41378\,f to an obliquity $\varepsilon\approx 90^\circ$, can also naturally provide the material for its hypothetical ring.

We stress, however, that even though this dynamical mechanism is physically realistic for HIP\,41378\,f, this does not imply that it necessarily happened. Planet HIP\,41378\,f may have had too small and/or too distant moons for the mechanism to operate, and the anomalous transit depth and flat spectrum of this planet may still be due to a particularly tenuous atmosphere covered with high-altitude hazes \citep{Chachan-etal_2020,Alam-etal_2022,Belkovski-etal_2022}. Yet, our analysis does provide further significance to the high-obliquity ring hypothesis, by showing that such an unusual configuration is not only feasible in a physical point of view, but even expected for some fraction of exoplanets resembling HIP\,41378\,f -- that is, for old and distant exoplanets in multi-planetary systems. As detailed above, checking the plausibility of this mechanism only requires a limited knowledge of the planetary system considered, and this methodology can be applied to other super-puff exoplanets, and in particular to the potential future discoveries of \emph{PLATO}.

\begin{acknowledgements}
The authors thank the anonymous referee for her/his inspiring comments. This work was supported by the Programme National de Plan{\'e}tologie (PNP) of CNRS/INSU, co-funded by CNES.
\end{acknowledgements}

\bibliographystyle{aa}
\bibliography{migrexomoon}

\appendix

\section{Correlations between parameters and orbital precession frequencies}\label{asec:correl}

In a long-term stable planetary system, the orbital motion of the planets can be approximated by quasi-periodic series as in Eq.~\eqref{eq:zeta}, where the frequencies $\nu_j$ are integer combinations of the fundamental frequencies of the system. In the Lagrange-Laplace approximation, the fundamental frequencies $s_k$ of the series governing the inclination dynamics of the planets are the eigenvalues of the matrix $B$ (see e.g. \citealp{Murray-Dermott_1999}) which only depends on the masses and semi-major axes of the planets.

\begin{figure*}
   \includegraphics[width=\textwidth]{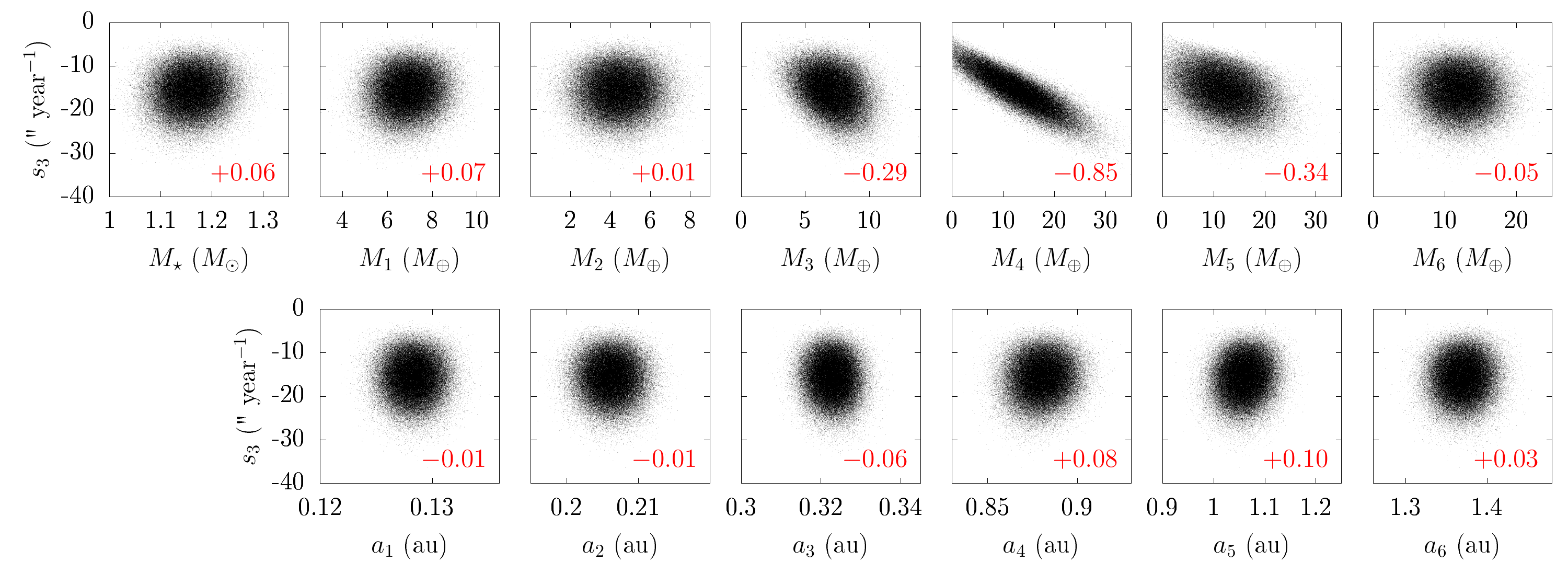}
   \caption{Scatter plot of the proper frequency $s_3$ and the 13 parameters involved (the stellar mass and the masses and semi-major axes of planets 1 to 6). The black dots show $150\,000$ realisations of the Lagrange-Laplace system with the mass and semi-major axis uncertainties from Table~\ref{tab:param}. The red label gives Spearman's correlation coefficient $\rho_\mathrm{S}$ as computed from $10^6$ realisations.}
   \label{fig:correl3}
\end{figure*}

\begin{figure*}
   \includegraphics[width=\textwidth]{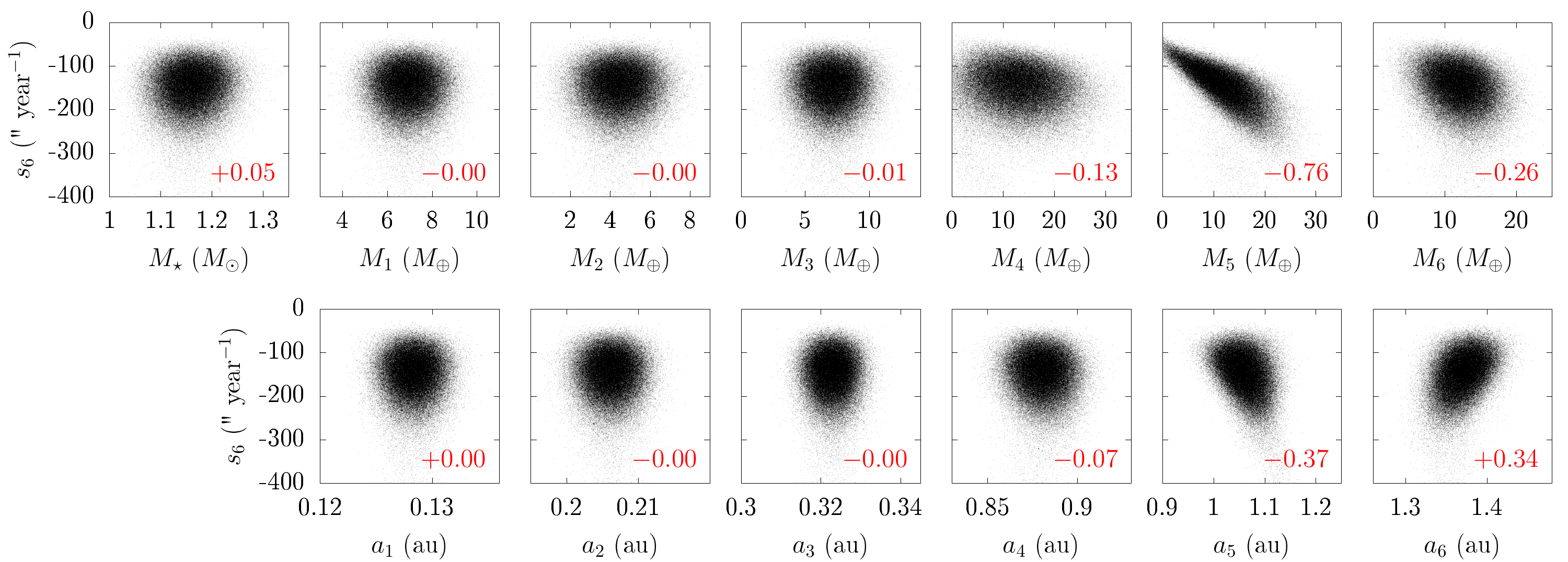}
   \caption{Same as Fig.~\ref{fig:correl3}, but for frequency $s_6$.}
   \label{fig:correl6}
\end{figure*}

The values of the frequencies $s_k$ and their role in the dynamics are intrinsic properties of the matrix $B$. As the fundamental frequencies reflect the gravitational couplings between the planets, some parameters contribute much more than others in the value of a given frequency; however, this contribution is not linear and it is not obvious a priori which parameters contribute the most. Here, we are mostly interested in the frequencies $s_3$ and $s_6$ as defined in Sect.~\ref{sec:orbit}, because they are expected to dominate the inclination dynamics of planet HIP\,41378\,f. Figures~\ref{fig:correl3} and \ref{fig:correl6} show the scatter of the values of $s_3$ and $s_6$ as a function of all parameters.

From these scatter plots, it is visible that the value of $s_3$ mostly depends on $M_4$, while the value of $s_6$ mostly depends on $M_5$. The strength of these correlations can be quantified by Spearman's correlation coefficient $\rho_\mathrm{S}$. The coefficients obtained are given in Figs.~\ref{fig:correl3} and \ref{fig:correl6} for each parameter. The correlations between $s_3$ and $M_4$ ($\rho_\mathrm{S}=-0.85$) and between $s_6$ and $M_5$ ($\rho_\mathrm{S}=-0.76$) are the strongest by more than a factor of two.

\section{Influence of the unknown physical parameters of the planet}\label{asec:param}

In Sect.~\ref{sec:moon}, we estimate the properties of a hypothetical former moon needed to tilt a planet from a low obliquity and create a ring of debris. The formulas, however, depend on the unknown parameters $J_2$ and $\omega\lambda$ of the planet. In this section, we consider them as free parameters and study their influence on our results.

\begin{figure}
   \includegraphics[width=\columnwidth]{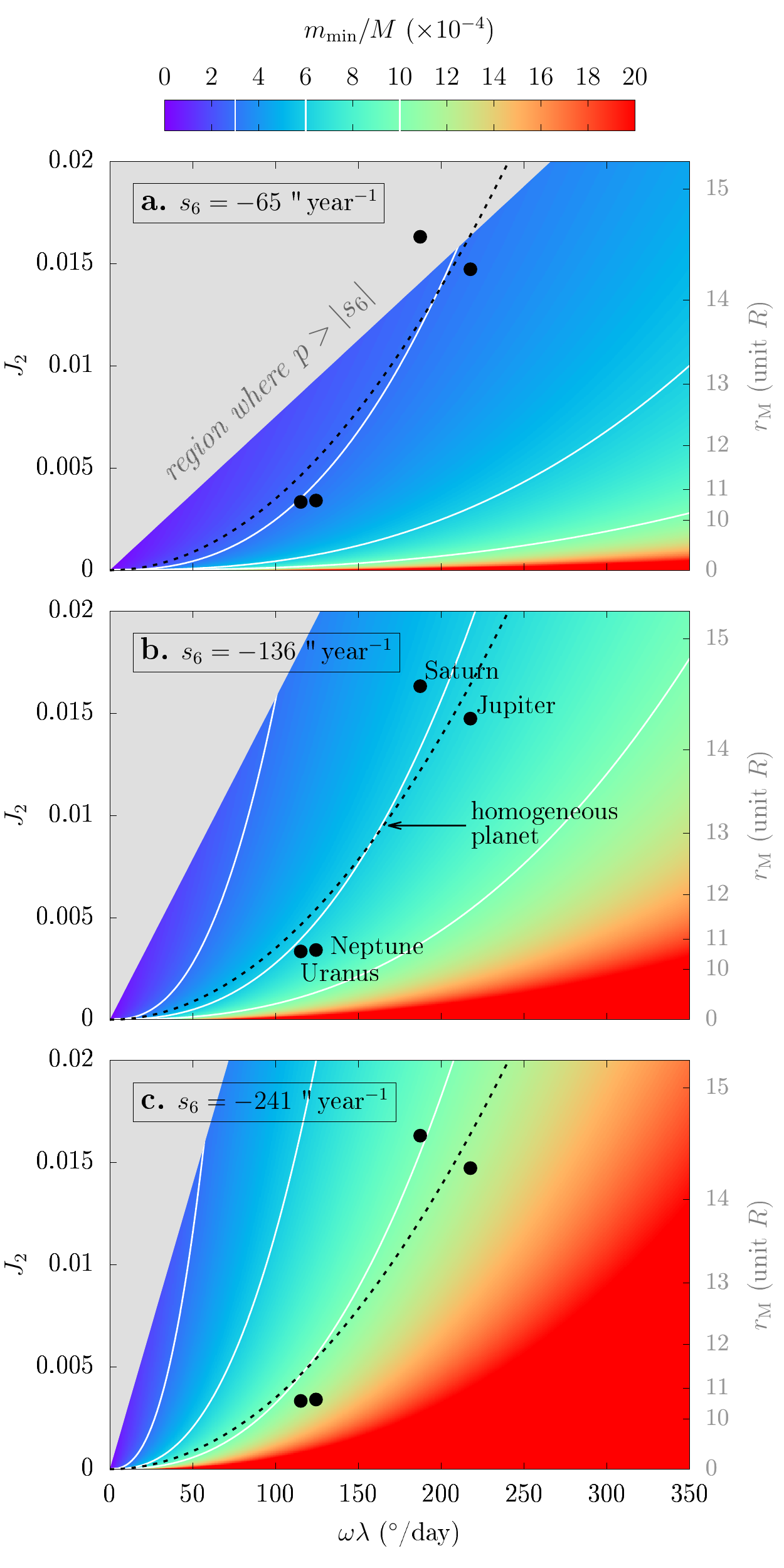}
   \caption{Minimum moon mass needed for HIP\,41378\,f to undergo the full tilting mechanism through a resonance with $s_6$. Each panel corresponds to a given possible value of $s_6$ (labels). The minimum moon mass $m_\mathrm{min}$ (colour scale) is obtained from Eq.~\eqref{eq:condtilt} as a function of the unknown parameters $J_2$ and $\omega\lambda$ of HIP\,41378\,f. The white curves highlight three mass levels, namely $m_\mathrm{min}/M = 3\times 10^{-4}$, $6\times 10^{-4}$, and $10\times 10^{-4}$. The grey region is forbidden by the left inequality in Eq.~\eqref{eq:condtilt}. The black dotted line shows the relation between $J_2$ and $\omega\lambda$ obtained for a homogeneous planet (see Eq.~\ref{eq:J2approx}). For reference, the black dots show the parameters $(\omega\lambda,J_2)$ of the Solar System giant planets \citep{Yoder_1995}.}
   \label{fig:mmin}
\end{figure}

The values of $J_2$ and $\omega\lambda$ are related and depend on the interior properties of the planet. In the simplest case of a homogeneous planet, $J_2$ and $\omega$ are linked through the law of Maclaurin's ellipsoid (see e.g. \citealp{Chandrasekhar_1969}), while $\lambda=2/5$. This law simplifies to $J_2\propto\omega^2$ for a nearly spherical planet, and it can be rewritten as
\begin{equation}\label{eq:J2approx}
  J_2 \approx \frac{25}{8}\frac{R^3}{\mathcal{G}M}(\omega\lambda)^2\,.
\end{equation}
Even though planets are not homogeneous, this approximate relation gives an idea of where to look for realistic combinations of parameters in the plane $(\omega\lambda,J_2)$ without any assumption on its composition. Figure~\ref{fig:mmin} shows that the Solar System giant planets do fall roughly along this curve.

For a given frequency $\nu_j$ in the orbital precession spectrum of a planet, the conditions $p\leqslant |\nu_j|$ in Eq.~\eqref{eq:condtilt} corresponds to a straight line $J_2\propto\omega\lambda$, whereas the condition $|\nu_j|\leqslant p\eta/2$ corresponds to a power law $J_2\propto(\omega\lambda)^{5/2}$. Both conditions can be visualised in Fig.~\ref{fig:mmin} for planet HIP\,41378\,f using three different values of the frequency $\nu_j$ (most probable value of $s_6$ and $95.4\%$ bounds; see Sect.~\ref{sec:orbit}). Because the exponent $2$ in Eq.~\eqref{eq:J2approx} is close to $5/2$, the realistic combinations of $J_2$ and $\omega\lambda$ (which are located in a rough neighbourhood of the dotted curve) follow more or less the level curves of the minimum moon mass $m_\mathrm{min}$. As a consequence, our estimate of the minimum moon mass is not affected much by our total ignorance of the parameters $J_2$ and $\omega\lambda$. Whatever realistic values are chosen, we obtain a minimum moon mass $m_\mathrm{min}/M\approx 6\times 10^{-4}$ for the most probable value of $s_6$ (Fig.~\ref{fig:mmin}b), with a dispersion at $95.4\%$ ranging from $m_\mathrm{min}/M\approx 3\times 10^{-4}$ (Fig.~\ref{fig:mmin}a) to $m_\mathrm{min}/M\approx 1\times 10^{-3}$ (Fig.~\ref{fig:mmin}c). These values are similar to those obtained in Sect.~\eqref{sec:moon} using the parameters of Uranus.

The value of characteristic length $r_\mathrm{M}$, however, depends on $J_2$, but not on $\omega\lambda$ (see Eq.~\ref{eq:rM}). The distance covered by the migrating moon depends therefore on the value chosen for $J_2$. Yet, this dependence is not very steep ($r_\mathrm{M}\propto J_2^{1/5}$). Moreover, extreme values of $J_2$, either very small or very large, can be ruled out because giant planets are expected to spin at a fraction of their breakup velocity (see e.g. \citealp{Batygin_2018,Dong-etal_2021,Dittmann_2021}). As a consequence, the value obtained for $r_\mathrm{M}$ only varies by a small amount when considering realistic values of $J_2$ (see the right vertical axis in Fig.~\ref{fig:mmin}).

\section{Inclination amplitudes and transit probabilities}\label{asec:amplitudes}

In the Lagrange-Laplace approximation, the long-term inclination dynamics of planets are described by quasi-periodic series as in Eq.~\eqref{eq:LLsol}, in which the frequencies solely depend on the masses and spacing of the planets. The orientations of the planets' orbital planes enter into play only in the amplitudes of the terms of the series. Apart from the zero-frequency term (which merely gives the orientation of the planetary invariant plane), the amplitudes $S_j$ in Eq.~\eqref{eq:LLsol} depend on the mutual inclinations between the planets' orbital planes. The mutual inclination $\Psi_{ik}$ of two planets $i$ and $k$ can be written as
\begin{equation}\label{eq:Imut}
   \cos\Psi_{ik} = \cos I_i\cos I_k + \sin I_i\sin I_k\cos (\Omega_i-\Omega_k)\,,
\end{equation}
where $I$ and $\Omega$ are the orbital inclination and longitude of ascending node of the planets measured with respect to a given reference plane (e.g. the sky plane). As a simple rule of thumb, the larger the mutual inclinations between each pair of planets, the larger the amplitudes $S_j$. For unknown longitudes of node $\Omega_k$, the orbital inclinations $I_k$ of a set of planets provide minimum and maximum bounds to the mutual inclinations between each pair of planets. From Eq.~\eqref{eq:Imut}, the minimum mutual inclination of two planets is $\Psi_{ik} = |I_i-I_k|$ reached for $\Omega_i-\Omega_k = 0$. In practice, inclination values measured from transit data have a mirror degeneracy with respect to $90^\circ$ (see Table~\ref{tab:param} of the main text); $\Psi_{ik}$ is minimised if $I_i$ and $I_k$ lie on the same side of $90^\circ$.

\begin{figure*}
   \includegraphics[width=\textwidth]{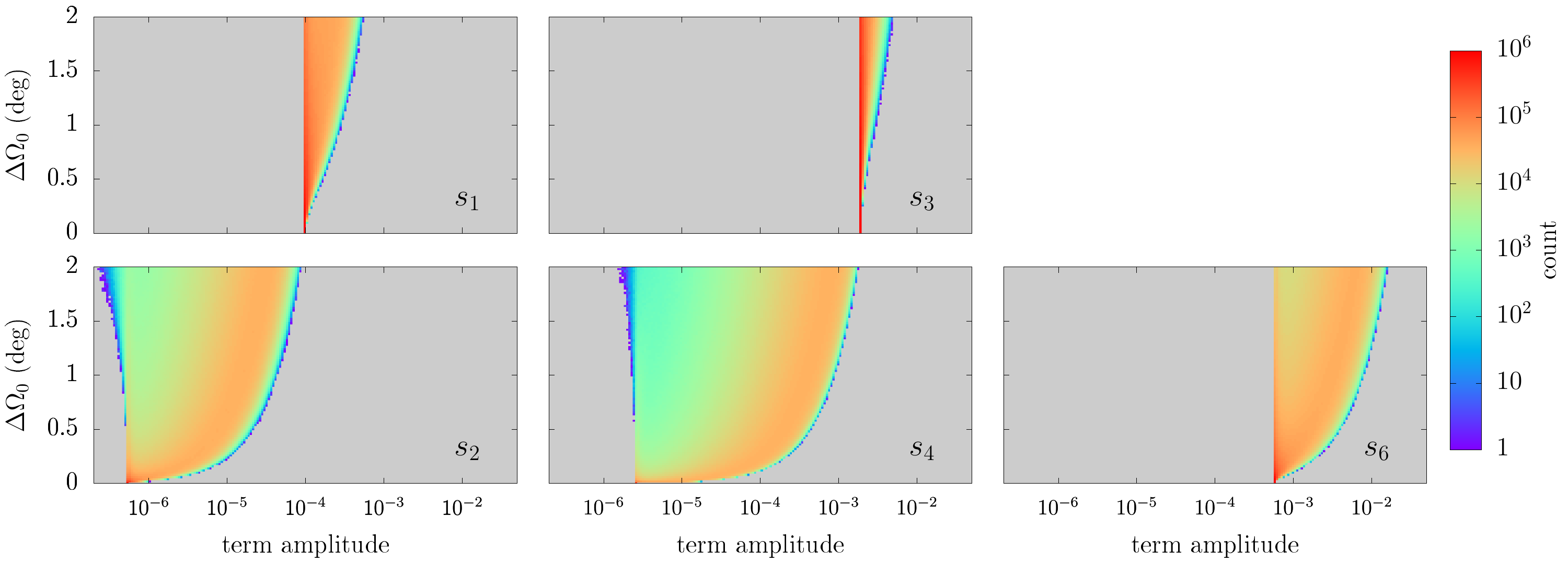}
   \caption{Possible values of the amplitudes $S_j$ for the inclination terms of planet HIP\,41378\,f. The frequency of each term is labelled. The amplitude of the zero-frequency term $s_5$ is not shown. The masses, semi-major axes, and inclinations of all planets are set to their nominal observed values (see Table~\ref{tab:param}); all inclinations values $I_k$ are assumed to be $I_k\leqslant 90^\circ$. The unknown longitudes of node $\Omega_k$ of the planets in the sky plane are drawn from a uniform random distribution in an interval $\Delta\Omega_0$ common for all planets. For a given value of the range $\Delta\Omega_0$ (vertical axis), a histogram of each amplitude is built from $10^6$ realisations of the Lagrange-Laplace system (colour scale). A bin is coloured grey if no occurrence is found among our $10^6$ realisations. The scale is the same for all graphs to highlight the disparity of amplitude between the different terms.}
   \label{fig:ampstats}
\end{figure*}

Here, we quantify the influence of the unknown longitudes of nodes on the orbital dynamics of planet HIP\,41378\,f by drawing randomly the longitudes of nodes of all planets in a given interval $\Delta\Omega_0$ and building a histogram of the amplitudes $S_j$ obtained. The result is shown in Fig.~\ref{fig:ampstats} for an interval $\Delta\Omega_0$ ranging from $0^\circ$ to $2^\circ$. As expected, the terms $s_1$ and $s_2$ have very small amplitudes; they contribute negligibly to the dynamics of planet HIP\,41378\,f. The dominant terms of the dynamics are $s_3$ and $s_6$, with the qualitative roles described in Sect.~\ref{sec:orbit}. The amplitudes of all terms are generally minimum for $\Delta\Omega_0=0$, and they cover a wider and wider range of possibilities when we allow the planets' longitudes of nodes to be substantially distinct. This effect is particularly visible for the $s_4$ term, for which a dispersion as small as $\Delta\Omega_0=1^\circ$ can make the amplitude increase by a factor $1000$ or so.

\begin{figure}
   \includegraphics[width=\columnwidth]{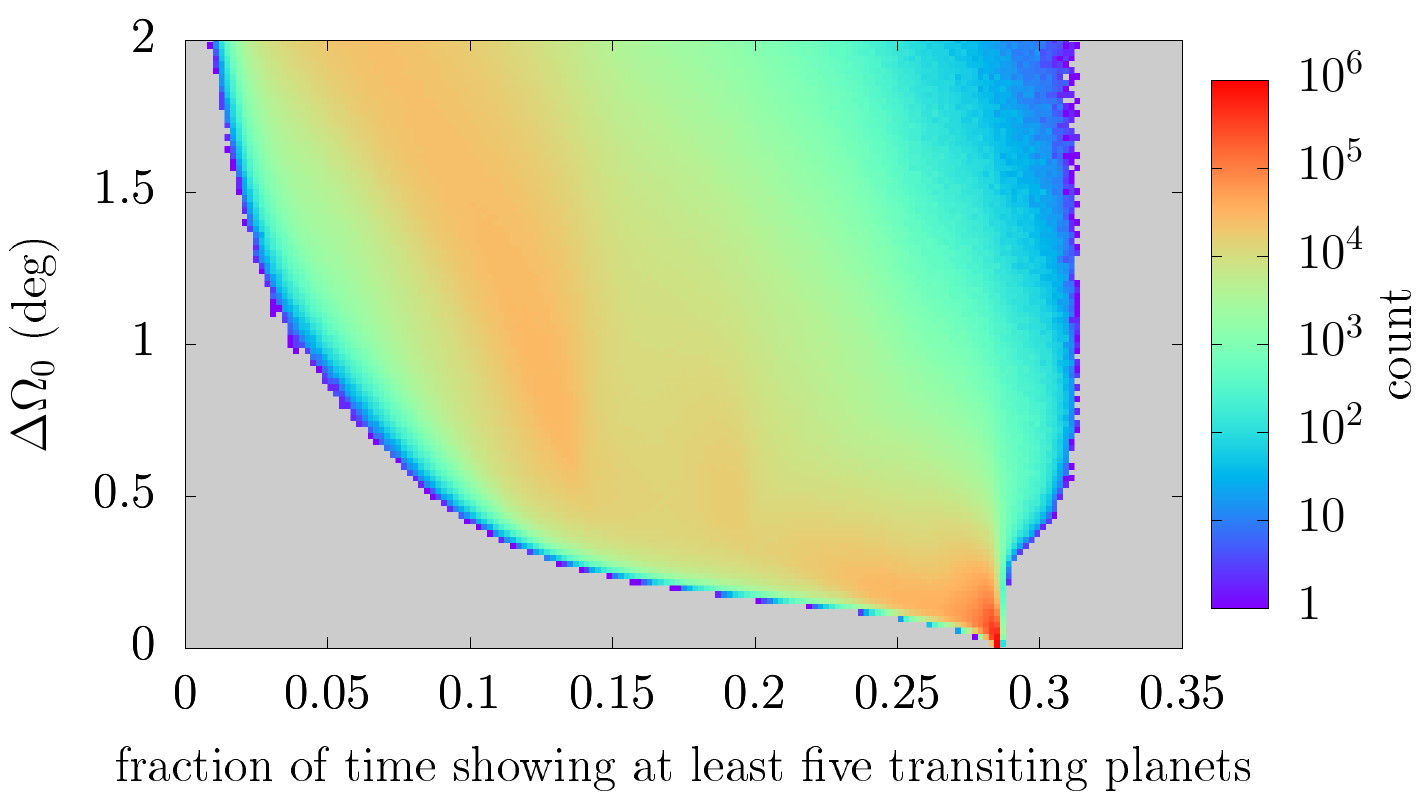}
   \caption{Probability of observing five transiting planets in the HIP\,41378 system over the planets' precession cycles. The unknown longitudes of node $\Omega_k$ of the planets in the sky plane are drawn from a uniform random distribution in an interval $\Delta\Omega_0$ common for all planets. For a given value of the range $\Delta\Omega_0$ (vertical axis), a histogram of the transit probability is built from $10^6$ realisations of the Lagrange-Laplace system (colour scale). A bin is coloured grey if no occurrence is found among our $10^6$ realisations.}
   \label{fig:transitprob}
\end{figure}

Figure~\ref{fig:transitprob} shows the probability of observing five transiting planets or more in the same experiment as in Fig.~\ref{fig:ampstats}. For a given realisation of the planetary system, the transit probability of a set of planets is the fraction of time their orbits simultaneously pass in front of the star. If we assume the same longitude of node for all planets in the HIP\,41378 system (i.e. $\Delta\Omega_0=0$), then five planets or more transit about $30\%$ of the time. If we allow a dispersion of $\Delta\Omega_0=1^\circ$, this fraction can be reduced to less than $5\%$.

\end{document}